\documentclass[reprint,amsmath,amssymb,prb,floatfix,twocolumn,superscriptaddress]{revtex4-1}

\usepackage{graphicx}
\usepackage{dcolumn}
\usepackage{bm}
\usepackage{times}
\usepackage[usenames,dvipsnames]{color}
\usepackage{colordvi,epsf}
\usepackage{braket}
\usepackage{float}
\usepackage{multirow}
\usepackage{comment}
\usepackage{longtable}
\usepackage[abs]{overpic}
\usepackage[utf8]{inputenc}
\usepackage[cmyk,dvipsnames]{xcolor}

\usepackage[all]{nowidow} 
\usepackage{microtype}

\begin{document}

\title{Effects of interlayer exchange on collapse mechanisms and stability of magnetic skyrmions}


\author{Hendrik Schrautzer}
\email[Email: ]{hes93@hi.is}
\affiliation{Institut f\"ur Theoretische Physik und Astrophysik,
Christian-Albrechts-Universit\"at zu Kiel, D-24098 Kiel, Germany}
\affiliation{Science Institute, University of Iceland, 107 Reykjavik, Iceland}
\author{Stephan von Malottki}
\affiliation{Institut f\"ur Theoretische Physik und Astrophysik,
Christian-Albrechts-Universit\"at zu Kiel, D-24098 Kiel, Germany}
\affiliation{Science Institute, University of Iceland, 107 Reykjavik, Iceland}
\author{Pavel F. Bessarab}
\affiliation{Science Institute, University of Iceland, 107 Reykjavik, Iceland}
\affiliation{ITMO University, 197101 St. Petersburg, Russia}
\author{Stefan Heinze}
\affiliation{Institut f\"ur Theoretische Physik und Astrophysik,
Christian-Albrechts-Universit\"at zu Kiel, D-24098 Kiel, Germany}

\date{\today}

\begin{abstract}
Theoretical calculations of thermally activated decay of skyrmions in systems comprising several magnetic monolayers are presented, with a special focus on bilayer systems. 
Mechanisms of skyrmion collapse are identified and corresponding energy barriers and thermal collapse rates are evaluated as functions of the interlayer exchange coupling and mutual stacking of the monolayers using transition state theory and an atomistic spin Hamiltonian. In order to contrast the results to monolayer systems, the magnetic interactions within each layer are chosen so as to mimic the well-established Pd/Fe/Ir(111) system.  
%
Even bilayer systems demonstrate a rich diversity of skyrmion collapse mechanisms that sometimes co-exist. For very weakly coupled layers, the skyrmions in each layer decay successively via radially-symmetric shrinking. Slightly larger coupling leads to an asymmetric chimera collapse stabilized by interlayer exchange. When the interlayer exchange coupling reaches a certain critical value, the skyrmions collapse simultaneously. 
Interestingly, the overall energy barrier for the skyrmion collapse does not always converge to a multiple of that for a monolayer system in the strongly coupled regime. For a certain stacking of the magnetic layers, the energy barrier as a function of the interlayer exchange coupling features a maximum and then decreases with the coupling strength in the strong coupling regime. 
Calculated mechanisms of skyrmion collapse are used to ultimately predict the skyrmion lifetime. Our results reveal a comprehensive picture of thermal stability of skyrmions in magnetic multilayers and provide a perspective for realizing skyrmions with controlled properties.
\end{abstract}

\maketitle

\section{Introduction}\label{chap:intro}
Over the past decade, topological spin textures such as magnetic skyrmions 
have been in the focus of many experimental and theoretical studies due to their intriguing properties \cite{back2020,Fert2017,wiesendanger2016}. 
After being predicted theoretically \cite{bogdanov1994}, the first experimental evidence of a skyrmion lattice was obtained in cubic B20 compounds\cite{muhlbauer2009,Yu2010}. 
The broken inversion symmetry in these crystals induces 
the Dzyaloshinskii-Moriya interaction (DMI)\cite{dzyaloshinsky1957,moriya1960} favoring noncollinear magnetic structures\cite{heide2008, perini2018}. %
Interfaces or surfaces naturally break the inversion symmetry, too, leading to interfacial DMI in ultrathin transition-metal films on substrates with significant spin-orbit coupling \cite{bode2007,ferriani2008}. %
This class of skyrmionic systems was established by the discovery of a nanoscale skyrmion lattice in monolayer Fe films on Ir(111)\cite{heinze2011}, and later enriched by experimental observation of skyrmions in ultrathin film systems such as Pd/Fe/Ir(111)~\cite{romming2013,romming2015}, 
Pd/Pd/Fe/Ir(111)~\cite{romming2019}, 3Fe/Ir(111)~\cite{hsu2017}, 
Co/Ru(0001)~\cite{herve2018}, and Rh/Co/Ir(111)~\cite{Meyer2019}. %

In ultrathin films, the magnetic interactions such as magnetic exchange, DMI and magnetocrystalline anisotropy can be tuned over a wide range via various mechanisms~\cite{parkin1991,ferriani2007,hardrat2009,blizak2012,Dupe2014,yang2015,Bellabes2016,beutier2017,yang2018,Meyer2019}, making these systems a convenient platform for realizing skyrmions with controlled properties\cite{juge2019}. %
%
%
Moreover, due to their pseudomorphic growth and the possibility of direct observation of their magnetic structures by surface-sensitive measurement techniques, ultrathin films became well-established model systems for the understanding of skyrmion properties\cite{heinze2011,romming2013,romming2015,grenz2017,herve2018,Meyer2019}. 

One major issue for the technological application of magnetic skyrmions is thermal stability, which is especially limited in ultrathin-films. %
%
Previous theoretical calculations applied to magnetic monolayers have predicted that a skyrmion state in the system coupled to the heat bath could decay into the topologically-trivial state via radially symmetric shrinking~\cite{lobanov2016,malottki2017,bessarab2015} or asymmetric collapse involving local rotation of magnetization at an excentric point of the skyrmion -- so called chimera mode~\cite{Meyer2019,desplat2019}. Both collapse modes have subsequently been discovered by means of spin-polarized scanning-tunneling microscopy in the Pd/Fe/Ir(111) system subject to an oblique external magnetic field~\cite{muckel2021}. Additionally, skyrmions are expected to be able to escape through the system's boundaries~\cite{bessarab2018} or even duplicate~\cite{muller2018}. 
The decay processes ultimately define the skyrmion lifetime, a quantitative measure of the skyrmion stability, which is usually described by an Arrhenius law \cite{bessarab2018,malottki2019,desplat2018}
\begin{equation}
\tau=\tau_0 \exp \left( \frac{\Delta E}{k_\mathrm{B} T} \right),
\label{Eq:arrhenius}
\end{equation}
where $\tau$ is the mean skyrmion lifetime, $\tau_0$ the pre-exponential factor, $\Delta E$ the energy barrier and $k_{\mathrm{B}} T$ the thermal energy. %

Recent atomistic simulations, either parameterized by \textit{first-principles}
density functional theory (DFT) calculations or as systematic parameter studies, revealed, that a large DMI, strong exchange frustration \cite{malottki2017}, the occurrence of higher order exchange interaction \cite{heinze2011,paul2020} or tuning of the skyrmion shape \cite{varentcova2020} can enhance skyrmion stability drastically. %
Furthermore, a decisive entropic stabilization effect has been found, increasing the prefactor of the Arrhenius law and thus, the skyrmion lifetime \cite{malottki2019,desplat2019,ritzmann2018,wild2017,varentcova2020}. %

Another theoretically predicted \cite{dupe2016} design strategy for improved skyrmion stability is the repeated stacking of additional magnetic layers, increasing the amount of magnetic material in the system. %
By sandwiching the magnetic layers between two different heavy metals, an additional enhancement of the effective DMI can be achieved as a result of additive interfacial chiral interactions, which additionally favors the stability of magnetic skyrmions \cite{MoreauLuchaire2016}. %
Indeed, by following the idea of multilayer systems, room-temperature stability of skyrmion has been achieved in different materials \cite{MoreauLuchaire2016,woo2016,boulle2016,soumyanarayanan2017}. %
In contrast to skyrmions in ultrathin film systems, however, skyrmions in multilayers have been found to be larger in size, typically on the order of $100$~nm \cite{MoreauLuchaire2016,woo2016,boulle2016,soumyanarayanan2017}. %
More recently, room-temperature skyrmions with sizes down to $30$~nm have been accomplished by using a compensated ferrimagnetic material \cite{caretta2018}. %
An additional advantage of multilayers compared to monolayer systems is the suppressed skyrmion Hall effect \cite{nagaosa2013} in antiferromagnetically coupled layers, as it hast been demonstrated by Legrand \textit{et al.} at room-temperature conditions and without external magnetic fields \cite{legrand2019}. Recently Rana \textit{et al.} also succeeded in stabilizing skyrmions at zero field at room temperature using the exchange-bias effect\cite{rana2020}. %

In contrast to the great success of its experimental realization, very little is understood about thermal stability of skyrmions in multilayer systems. %
In 2017, Stosic \textit{et al.} \cite{stosic2017} investigated the stability and collapse mechanisms of skyrmions in trilayers, focusing on the variation of DMI in the different layers. They showed that magnetic interactions differ significantly in a multilayer structure with varying thickness of the magnetic material due to the different interfaces the individual magnetic layers experience. The layer resolved and thus reduced DMI led to more realistic but less stable skyrmions than previously considered.
More recently, Hoffmann \textit{et al.} found an increasing skyrmion stability for an increasing number of magnetic layers. They assumed similar magnetic properties in each layer, a strong interlayer exchange coupling and a simultaneous radial symmetric collapse of skyrmions in all layers\cite{hoffmann2020}. Consistent with these general assumptions, Heil \textit{et al.} suggested in 2019 that the energy barrier for skyrmions in such systems is a multiple of the energy barrier of skyrmion collapse in the corresponding monolayer system\cite{Heil2019}, which reads
\begin{equation}
\Delta E\, =\, L\, \Delta E_{\text{mono}},   
\label{Eq:l_emono_conjecture}
\end{equation}
where $L$ is the number of stacked layers and $\Delta E_{\text{mono}}$ the energy barrier of the monolayer system. %
%

In this work, we systematically study the role of the interlayer exchange for skyrmion stability and the different regimes and effects it induces. %
For this purpose, we investigate bilayer and multilayer systems consisting of an artificial repetition of the famous Pd/Fe/Ir(111)\cite{malottki2017,malottki2019,Bottcher2018,Dupe2014, hagemeister2015,hanneken2015,leonov2016,romming2013,romming2015,rosza2016,simon2014} monolayer system. %
Since Dup\'e \textit{et al.} \cite{dupe2016} showed 
based on DFT calculations
that the magnetic interactions are primarily affected by the interfaces of the magnetic material, one can expect the properties of the magnetic layers in such a stacking to be comparable to the monolayer system. %
In order to obtain a broader view of the emerging effects, we vary the strength of the interlayer exchange coupling, $J^\perp$ systematically from zero to $20$~meV, coping indirectly and weakly coupled to directly and strongly coupled systems. %
Further, we explore two different crystal structures of the multilayer-stackings, revealing an exchange-bias-like effect in fcc and hcp structured systems, strongly affecting skyrmion stability.

The paper is structured as follows: Sec.~\ref{chap:model} describes the model and Sec.~\ref{chap:method} the method 
and computational details of our calculations. The presentation of our results in Sec.~\ref{Kap: Ergebnisse} starts with a brief discussion of the phase diagram (Sec.~\ref{ssec:phase_dia}) for magnetic bilayer systems under the influence of interlayer exchange. In Sec.~\ref{ssec:introduction_to_mechanisms} we discuss collapse mechanisms of skyrmions in magnetic bilayers into the field-polarized state, increasing the interlayer exchange stepwise and analyzing the occurring changes of the collapse mechanism. These results are subsequently condensed in Sec.~\ref{ssec:barriers} by studying the corresponding energy barriers. In Sec.~\ref{ssec:stacks_effective_model} we explain a crossover between two collapse mechanisms for critical interlayer exchange couplings. To understand these critical parameters in more detail, we then vary the DMI and hence the energy barrier of skyrmions in the underlying monolayer system in Sec.~\ref{ssec:vary_monolayer_barrier}. Afterwards we demonstrate that our results transfer to systems with more than two magnetic layers in Sec.~\ref{ssec:beyond2lay}. Finally, in Sec.~\ref{ssec:lifetime} we discuss calculations of the lifetime of bilayer skyrmions for a generic example. In Sec.~\ref{chap:conclusion}, we briefly conclude.

\begin{figure}
\includegraphics[scale=1]{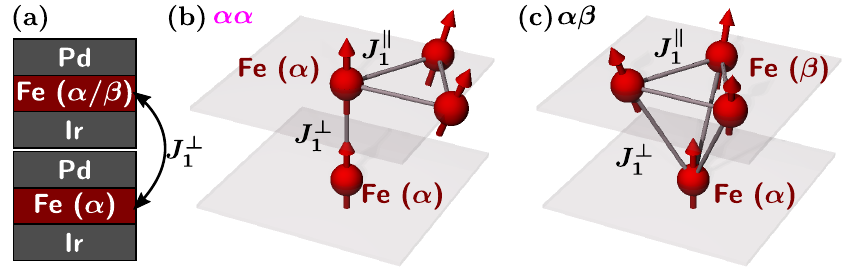}
\caption{(a) Magnetic bilayer systems built from two units of the system Pd/Fe/Ir(111). Two different stackings ($\alpha\alpha$ and $\alpha\beta$) are considered which correspond to atoms of the two hexagonal Fe
layers being on top of each other or shifted with respect to each other
as in fcc stacking, respectively. (b,c) Schematic representation of the
nearest-neighbor
intralayer ($J_1^\parallel$) and interlayer ($J_1^\perp$) exchange bounds for the $\alpha\alpha$-stacked and $\alpha\beta$-stacked magnetic bilayer, respectively. The bonds are indicated through the connections of the red magnetic moments.}
\label{fig:model_stacking}
\end{figure}
\section{Model}\label{chap:model}
The model for our spin simulations is shown in Fig.~\ref{fig:model_stacking}. We treat different 
stacking possibilities of the magnetic monolayer system Pd/Fe/Ir. 
Note, that only the hexagonal Fe layers of the system are included in our
atomistic spin model.
The effect of the nonmagnetic Pd and Ir layers is included within the framework of the first-principles parametrization of the 
magnetic interactions given 
in Ref.\cite{malottki2017} for the magnetic monolayer system (cf.~Sec.~\ref{chap:heisenberg}). Two different bilayers were studied. The system in which the magnetic moments in both Fe layers occupy the same lattice sites is called $\alpha\alpha$-system in the following (Fig. \ref{fig:model_stacking}(b)). In contrast, the magnetic moments of the $\alpha\beta$-system occupy the lattice sites of an fcc- or hcp-stacking of the Fe layers 
(Fig.~\ref{fig:model_stacking}(c)). 

We systematically vary the strength of the interlayer exchange between the Fe layers in our simulations. 
Therefore, the obtained results can be applied to systems in the
strong interlayer exchange coupling regime such as 
directly adjacent Fe layers, e.g.~in the system Rh/Pd/2Fe/2Ir\cite{dupe2016},
as well as in the weak or intermediate regime such as magnetic layers in which 
the interlayer exchange is mediated by a number of spacer layers.

\subsection{Extended Heisenberg model}\label{chap:heisenberg}
The magnetic bilayer systems are built based on the magnetic interactions of the monolayer system Pd/Fe/Ir(111) and described through normalized magnetic moments $\mathbf{m}_i$ localized in each Fe layer at the sites of a hexagonal lattice. 
The energy of the $N$-spin system is derived within the extended Heisenberg model and the Hamiltonian can be written as
\begin{align}
H=&E_\text{ex} + E_\text{DMI} + E_\text{Ani} + E_\text{Zee}\notag\\
=&-\sum\limits_{\substack{i,j=1 \\ i\neq j}}^N J_{ij} (\mathbf{m}_i\cdot\mathbf{m}_j)-\sum\limits_{\substack{i,j=1 \\ i\neq j}}^N \mathbf{D}_{ij}\cdot(\mathbf{m}_i\times\mathbf{m}_j)\notag\\
-&\sum\limits_{i=1}^N K (m_i^z)^2-\sum\limits_{i=1}^N\mu_i (\mathbf{m}_i\cdot\mathbf{B}_\text{ext}),
\label{gl:energy_model}
\end{align}
which are in the order of appearance the Heisenberg exchange, the DMI, the uniaxial magnetocrystalline anisotropy and the Zeeman interaction. The interaction constants $J_{ij}$, $\mathbf{D}_{ij}$ and the anisotropy constant $K$ are defined per atom. Therefore each pair of magnetic moments appears twice in the calculation of the exchange and DMI energy. Note, that we consider intralayer DMI here, but not interlayer DMI~\cite{vedmedenko2019,han2019}.

The exchange term can be split into intralayer exchange $J_{ij}^{\parallel}$ and interlayer exchange $J_{ij}^\perp$ for pairs of magnetic moments from the same layer and from different layers, respectively.
\begin{align}
    E_\text{ex} &= E_\text{ex}^{\parallel} + E_\text{ex}^{\perp}\notag\\
    &= -\sum\limits_{l=1}^{2}\sum\limits_{\substack{i,j=1 \\ i\neq j}}^{N_l}J_{ij}^{\parallel}(\mathbf{m}_i^l\cdot\mathbf{m}_j^l) 
    -\sum\limits_{i,j=1}^{N_1, N_2}J_{ij}^{\perp}(\mathbf{m}_i^1\cdot\mathbf{m}_j^2)
\end{align}
Here $N_l$ denotes the number of spins in the layer $l$.

Motivated by the finding of Dupé \textit{et al.}\cite{dupe2016} 
that the magnetic interactions in multilayers built from Pd/Fe/Ir stacks are very similar to
those of the film system Pd/Fe/Ir(111)
all intralayer interaction constants, i.e.~within a single Fe layer,
and the magnetic moments $\mu_i$ 
were taken from Pd/Fe/Ir(111)~\cite{malottki2017} as obtained via DFT 
calculations using the {\tt FLEUR} code
\cite{Dupe2014,Kurz2004,Heide2009,Zimmermann2014}. 
In Ref.~\onlinecite{malottki2017}
two different models were used to illustrate the influence of intralayer exchange frustration. On one hand, exchange constants were determined from DFT up to the interaction of ninth neighbors ($J_1^{\parallel}$,$\dots$,$J_9^{\parallel}$). We will refer to this set of parameters as the neighbor resolved exchange (NRE) model. On the other hand, the magnetic interactions of the system were parameterized with only the nearest-neighbor exchange interaction, which 
resembles a micromagnetic description of the interactions. The resulting 
parameter set is referred to as the effective model. %
The values of all parameters used in this work are listed in Tab.~\ref{tab: Pd_Fe_Ir_parameters} in the Appendix. 

We treat the interlayer exchange coupling in our magnetic Fe bilayers
in nearest-neighbor approximation
and systematically vary its strength, $J_{1}^\perp$.
As visible in Fig.~\ref{fig:model_stacking} the magnetic unit cell of the $\alpha\beta$-system contains three interlayer bonds while in the unit cell of the $\alpha\alpha$-system only one bond appears. We define the interlayer exchange per unit cell $J^{\perp}$ for better comparability of the different systems as the following:
\begin{align}
J^\perp=\begin{cases}J_1^\perp \text{,}&\alpha\alpha\text{-system}\\3\cdot J_1^\perp \text{,}&\alpha\beta\text{-system}\end{cases}.
\end{align}

\section{Computational details}\label{chap:method}
We use atomistic spin dynamics simulations to solve the Landau-Lifshitz-equations
for the spin model introduced in the previous section
numerically and to relax spin structures such as bilayer skyrmions into local energy minima. The knowledge of the separating energy barrier $\Delta E$ between meta-stable spin structures on the energy surface is crucial for the description of the thermal stability of these states following an Arrhenius law for the skyrmion lifetime $\tau$ (Eq.~(\ref{Eq:arrhenius})).
The geodesic nudged elastic band method\cite{bessarab2015}(GNEB) provides a possibility to calculate the energy barrier and the first-order saddle point of skyrmions regarding a transition to the topologically trivial ferromagnetic state.  We use the harmonic approximation of the transition-state theory (HTST) for determining the Arrhenius pre-exponential factors and the lifetimes of magnetic states\cite{bessarab2012}. While the phase diagrams presented in Sec.~\ref{ssec:phase_dia} are calculated with simulation boxes of $100~\times~100$ magnetic atoms per layer, all other results of this work are obtained with boxes of $50~\times~50$ magnetic atoms per layer. We applied periodic boundary conditions in in-plane direction, while open boundaries are assumed in out-of-plane direction.
Consistency tests for $70~\times~70$ and $100~\times~100$ magnetic moments per layer demonstrated that the shown results are not dependent on the system size.


\begin{figure}
\begin{overpic}[scale=1]
{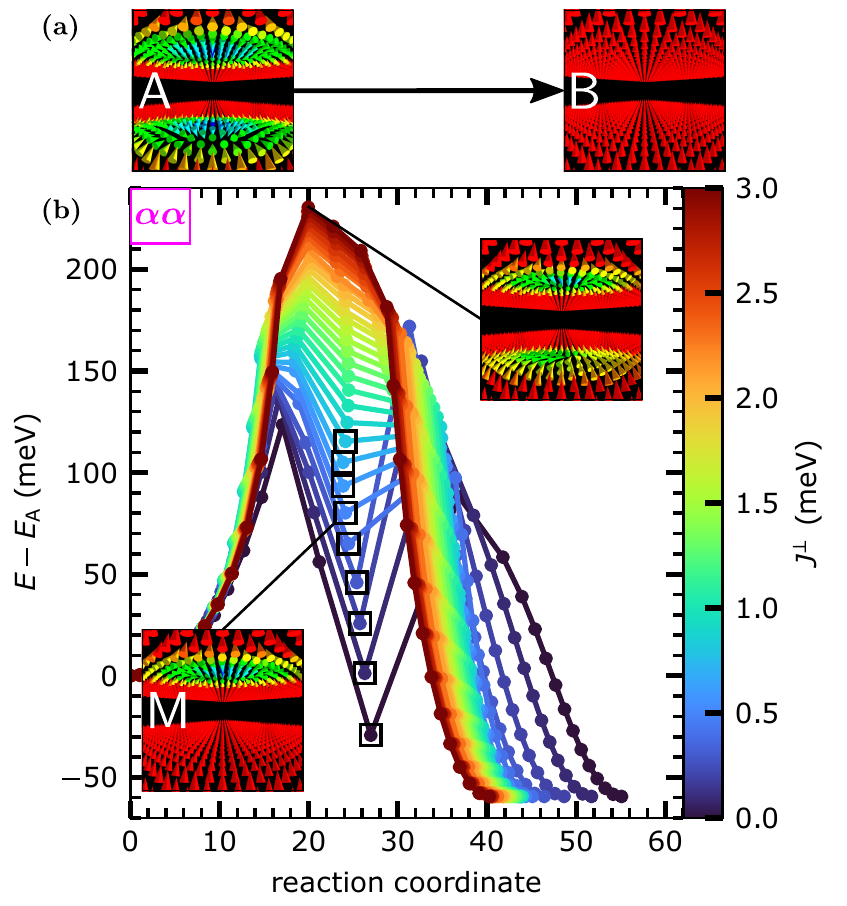}
\end{overpic}
\caption{(a) Representation of a bilayer skyrmion (initial state, A) collapsing into the field polarized state (final state, B). (b) Partially relaxed energy paths for the collapse mechanism of a bilayer skyrmion (A) for the $\alpha\alpha$-system for various interlayer exchange couplings $J^\perp$ (visualized by the color code). The shown paths are the results of 500 iterations of a GNEB calculation an do not display the converged minimum energy path. They are the starting point for treating paths with an intermediate energy minimum (M), which are marked with the empty squares. The insets show spin configurations for two interlayer exchange couplings. For $J^\perp=3.0$~meV the configuration with the highest energy is shown while for $J^\perp=0.5$~meV the spin configuration of the intermediate minimum is displayed.
}
\label{fig:GNEB_method}
\end{figure}
\subsection{Minimum energy path calculations}\label{ssec:method_GNEB}
The GNEB method is a valuable approach\cite{malottki2017,malottki2019,bessarab2018,bessarab2015, stosic2017, Rybakov2015} to calculate the minimum energy path (MEP) between magnetic configurations corresponding to local energy minima. 
As schematically illustrated in Fig.~\ref{fig:GNEB_method}(a) we consider the collapse 
of an initial magnetic state (A), which is a bilayer skyrmion, to the final magnetic state (B), which is the ferromagnetic or field polarized state.
In Sec.~\ref{Kap: Ergebnisse} we 
discuss the occurrence of different collapse mechanisms and the associated MEPs caused by the variation of the interlayer exchange. 
For 
weak interlayer exchange, paths with additional local minima between the initial and final states occur 
(cf.~Fig.~\ref{fig:GNEB_method}(a)). 
These intermediate minima (M) are associated with a successive collapse of the skyrmion in the different layers. We split up the paths at the states M after short GNEB calculations (500 iterations), as suggested in Ref. \cite{bessarab2015}. %
The energies of these partially relaxed paths are visualized on the example of bilayer skyrmions in the $\alpha\alpha$ system in Fig.~\ref{fig:GNEB_method}(b).
After the splitting, the M configuration is relaxed into its local energy minimum via spin dynamics. Afterwards, we calculate the MEPs for  $A\rightarrow M$ and $M\rightarrow B$ transitions separately with the GNEB method and finally connect them to create the complete paths $A\rightarrow B$. Consequently, there are paths with two first order saddle points (Sp) ($\text{Sp}_1$, $\text{Sp}_2$) for low values of the interlayer exchange couplings and paths with one saddle point for strong interlayer exchange couplings. These saddle points are determined with the climbing-image GNEB method (CI-GNEB)\cite{bessarab2015}. A calculation is considered converged when the force on each magnetic moment has dropped below  $10^{-8}$~eV/rad.

\subsection{Harmonic transition-state theory}\label{ssec:method_HTST}
We determine the pre-exponential factor $\tau_0$ within the harmonic approximation of the TST. This implies the description of the curvature of the multidimensional energy surface of the spin configuration room via the eigenvalues $\epsilon_{\text{A},i}$ and $\epsilon_{\text{Sp},i}$ of the Hessian matrices $H_\text{A}$ and $H_\text{Sp}$ for the bilayer skyrmion configuration (A) and the saddle point configuration (Sp), respectively. In the general form the pre-exponential factor is given by\cite{bessarab2018,bessarab_uzdin2012, stephan_phd}
\begin{align}
    \tau_0^{-1}=\frac{\lambda}{2\pi}\left(2\pi k_B T\right)^{(P_\text{A}-P_\text{Sp})/2}\frac{V_{\text{Sp}}}{V_\text{A}}\sqrt{\frac{\operatorname{det}H_{\text{A}}}{\operatorname{det}'H_\text{Sp}}}.
    \label{gl:HTST}
\end{align}
The determinants of the Hessian matrices at the bilayer skyrmion and the saddle point state are computed as the product of the corresponding nonzero eigenvalues, while the prime indicates that the negative eigenvalue for the saddle point is omitted. The information about the velocity of the system at the transition state is contained by the dynamical factor $\lambda$ (see Ref.~\cite{varentcova2020} for details).
Not all eigenmodes are suited for a description in harmonic approximation. Alternatively, Goldstone modes can be defined and calculated as such \cite{bessarab2018}, yielding the Goldstone mode volumes $V_\text{A}$, $V_\text{Sp}$, while the corresponding eigenvalues are omitted in the determinants of Eq.~(\ref{gl:HTST}) as well. The number of Goldstone mode for the initial state (saddle point) is given by $P_\text{A}$ ($P_\text{Sp}$). %
In the case of skyrmion annihilation in bilayers, this applies to the two skyrmion translation modes in in-plane direction as the movement of skyrmions over the lattice does not change their energy, similar to the translation of skyrmions in monolayer systems \cite{malottki2019,varentcova2020}. %
Throughout this work, we investigate the skyrmion lifetime only in cases, in which the simultaneous radial symmetric collapse mechanism \cite{malottki2017,bessarab2018,muller2018, hoffmann2020} is dominant for the annihilation process. %
The corresponding saddle point structures contain three neighboring magnetic moments pointing almost towards each other, creating a Bloch-like point (see Fig. \ref{fig:overview_collapses})~(p). %
For low temperatures, this Bloch-like point cannot be moved without noticeable energy costs over the atomic lattice. %
For elevated temperatures, the eigenmodes corresponding to this movement are potential Goldstone modes in the spectrum of the saddle point state, as discussed in Ref. \cite{varentcova2020,desplat2019}. %
For the sake of clarity, however, here we treat all eigenmodes of the saddle point structures in harmonic approximation and thereby exclude the high temperature regime. %

The unequal number of Goldstone modes found for the skyrmion and saddle point states leads to a linear temperature dependence for the inverse of the pre-exponential factor\cite{malottki2019}:
\begin{align}
    \tau_0^{-1}=\frac{2\lambda k_B T}{V_\text{A}}\sqrt{\frac{\operatorname{det}H_{\text{A}}}{\operatorname{det}'H_\text{Sp}}}.
    \label{eq:prefactor_radial}
\end{align}
The factor of two arises from the two possible realizations of the Bloch-like point per unit cell, as discussed in Ref.~\cite{malottki2019}. 

\section{Results}\label{Kap: Ergebnisse}

\subsection{Zero temperature phase diagrams}\label{ssec:phase_dia}
To study the metastability of skyrmions in the field-polarized phase, first we have to determine the critical fields 
which correspond to 
the phase transition between the skyrmion lattice and the field-polarized phase. %

Therefore, obtaining the zero temperature magnetic phase diagrams\cite{Dupe2014,malottki2017,Bottcher2018} of the $\alpha\alpha$- and $\alpha\beta$-system as a function of interlayer exchange coupling is the starting point for our investigations. In the following, we present the phase diagrams calculated within the NRE~model of intralayer exchange interaction. 

\begin{figure}
\begin{overpic}[scale=1]
{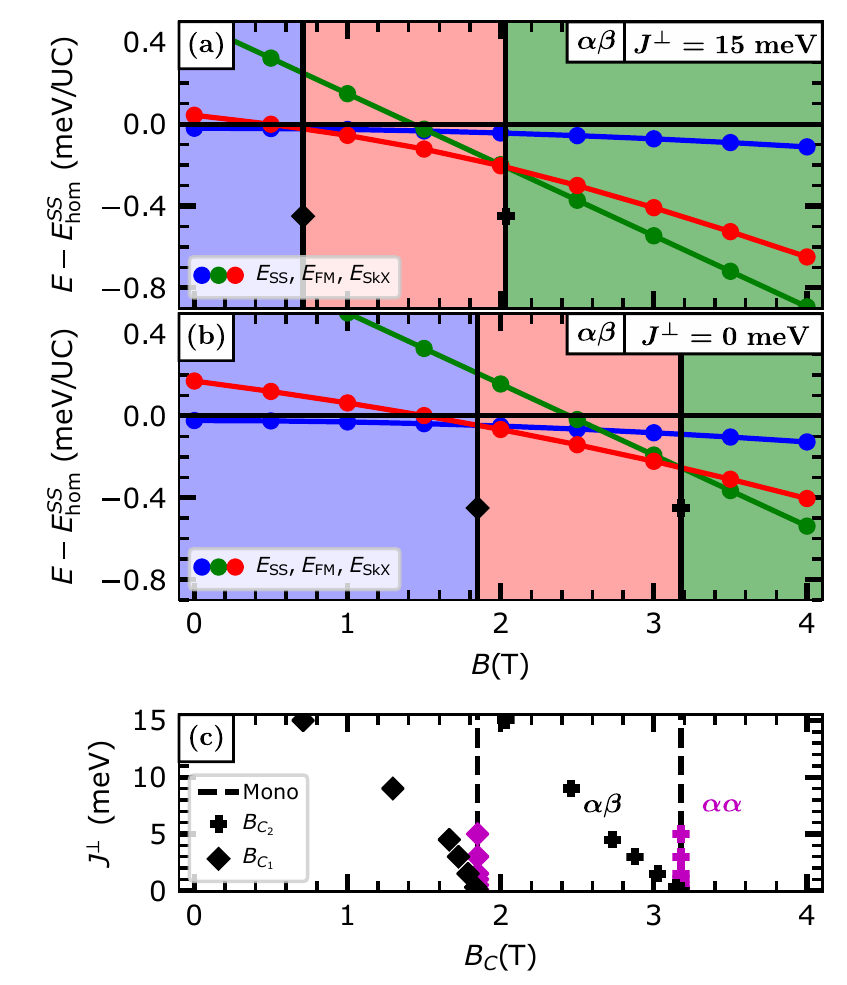}
\end{overpic}
\caption{(a) Zero temperature phase diagram for the $\alpha\beta$ system for $J^\perp=15.0$~meV. The critical fields $B_{C_1}$ and $B_{C_2}$ define the phase transitions between the spin spiral phase (SS, blue), the skyrmion lattice phase (SkX, red) and the field polarized phase (FM, green), respectively. The energy is defined per unit cell and displayed relative to the energy of the minimum of the dispersion of homogeneous spin spirals.
The background color represents the phases in a certain magnetic field range. (b) Analog visualization of the phase diagram for the $\alpha\beta$ system for $J^\perp=0$~meV. (c) Critical magnetic fields $B_{C_1}$ or $B_{C_2}$ for different values of $J^\perp$ for the $\alpha\alpha$ and $\alpha\beta$ system. The reference of the magnetic monolayer system Pd/Fe/Ir(111) is plotted as a dashed line.}
\label{fig:phasedia}
\end{figure}

In Fig.~\ref{fig:phasedia}~(b), the energies of relaxed bilayer spin spirals (SS), bilayer skyrmion lattices (SkX) and the field polarized phase (FM) are shown over varying magnetic field strength for the $\alpha\beta$-system without interlayer coupling ($J^\perp=0$~meV). %
Similar to Ref.~\cite{malottki2017} we chose the energy reference as the minimum energy of the dispersion of homogeneous spin spirals,
$E_{\text{hom, SS}}$, 
calculated in a $100\,\times\,100$ simulation box. 
Further, we consider the 
SkX~state 
with the energetically most favorable skyrmion density on the $100\,\times\,100$ lattice.
The critical magnetic field values $B_{C_1}$ and $B_{C_2}$ mark the phase transitions from the SS~state to the SkX~state and from the SkX~state to the FM~state, respectively. 
As the energy is defined per unit cell and the interlayer exchange is switched off, these critical fields exactly coincide with the fields reported in Ref.~\cite{malottki2017} for the magnetic monolayer system Pd/Fe/Ir(111).

When we increase the interlayer exchange to $J^\perp=15$~meV for the $\alpha\beta$-system (See Fig.~\ref{fig:phasedia}(a)) the critical fields $B_{C_1}$ and $B_{C_2}$ shift to lower fields and thereby introduce a shift of the SkX phase. The origins of these energy shifts can be understand by considering the horizontal displacements between the layers (See Fig.~\ref{fig:model_stacking}(c)). A parallel alignment of two magnetic moments in different layers leads to a minimal exchange energy for ferromagnetic interlayer exchange. Therefore, the FM state
gains more energy than the SS
state, in which small angles between the magnetic moments
of adjacent Fe atoms in the two layers occur that 
are unfavorable with respect to the interlayer exchange.
These angles arise due to the horizontal displacement of the magnetic layers in the $\alpha\beta$-system. The SkX-phase lies between those two extremes as there are collinear aligned regions between the skyrmions
in the two layers
and therefore its energy shift is smaller than for the FM
state but greater than for the SS state
which leads to a decrease of both $B_{C_1}$ and $B_{C_2}$. Fig.~\ref{fig:phasedia}(c) underlines this behavior as it displays the decrease of the critical fields with increasing interlayer exchange $J^\perp$. In addition, it is noteworthy that the skyrmion density of the SkX~phase is slightly reduced for high interlayer exchange couplings. These effects can be summarized in the observation that even small angles between the magnetic moments of interacting magnetic layers lead to an exchange bias effect mediated by interlayer exchange\cite{chen2015,nandy2016}.%
 In contrast, for $\alpha\alpha$ systems for each magnetic moment the next neighbor regarding the interlayer exchange coupling is directly above or below the corresponding moment. Therefore the magnetic moment and its neighbor are aligned parallel for each magnetic structure considered in the phase diagram and the shifts in the energy are equal when varying the interlayer exchange. Fig.~\ref{fig:phasedia}~(c) presents these results by visualizing the critical fields $B_{C_1}$ and $B_{C_2}$. The dashed lines indicate the corresponding fields as determined for the magnetic monolayer system Pd/Fe/Ir(111)\cite{malottki2017}. Therefore the phase diagram remains unchanged for $\alpha\alpha$-systems when varying the interlayer exchange and this will also hold true for systems with more magnetic layers if the atoms of each layer occupy the same lattice sites.

\subsection{Skyrmion collapse mechanisms in magnetic bilayers}
\label{ssec:introduction_to_mechanisms}
For a detailed understanding of the thermal stability of magnetic bilayer skyrmions (A) in the field polarized phase the MEP regarding a collapse to the ferromagnetic aligned structure (B) is crucial 
(Fig.~\ref{fig:GNEB_method}). 
To be consistent with the calculations of the underlying monolayer\cite{malottki2017} system we chose an out of plane magnetic field of $B=4.0$~T. It has be shown that the skyrmion sizes for the effective and the NRE model are similar for $B=4.0$~T\cite{malottki2017}, which allows a reasonable comparison between the NRE and the effective model. Our phase diagram calculations in the previous sections demonstrated that $B_C(J^\perp)<B$ holds true for all interlayer exchange couplings for both stackings of the system (cf.~Fig.~\ref{fig:phasedia}(c)). 
Therefore, we expect isolated bilayer skyrmions to be meta-stable in both systems at $B=4.0$~T. Note, that in the case of the $\alpha\beta$ system the distance $B-B_C(J^\perp)$ increases with increasing interlayer exchange. A decreased stability of skyrmions in the $\alpha\beta$ system with increasing interlayer exchange can be expected, as elucidated in Sec.~\ref{ssec:barriers} in more detail. In that context it is worth mentioning that the skyrmion radius of the bilayer skyrmions in the $\alpha\beta$-system is marginally reduced when increasing $J^\perp$, which follows the relation between skyrmion size and stability\cite{varentcova2018}. For the highest values of $J^\perp$ in our work the reduction in the skyrmion size is less than one in-plane lattice constant. However, the radius of the bilayer skyrmions in the $\alpha\alpha$-system agrees for all values of $J^\perp$ with the radius reported for monolayer skyrmions in Pd/Fe/Ir(111)\cite{malottki2017}.

\begin{figure}
\begin{overpic}[scale=1.0]
{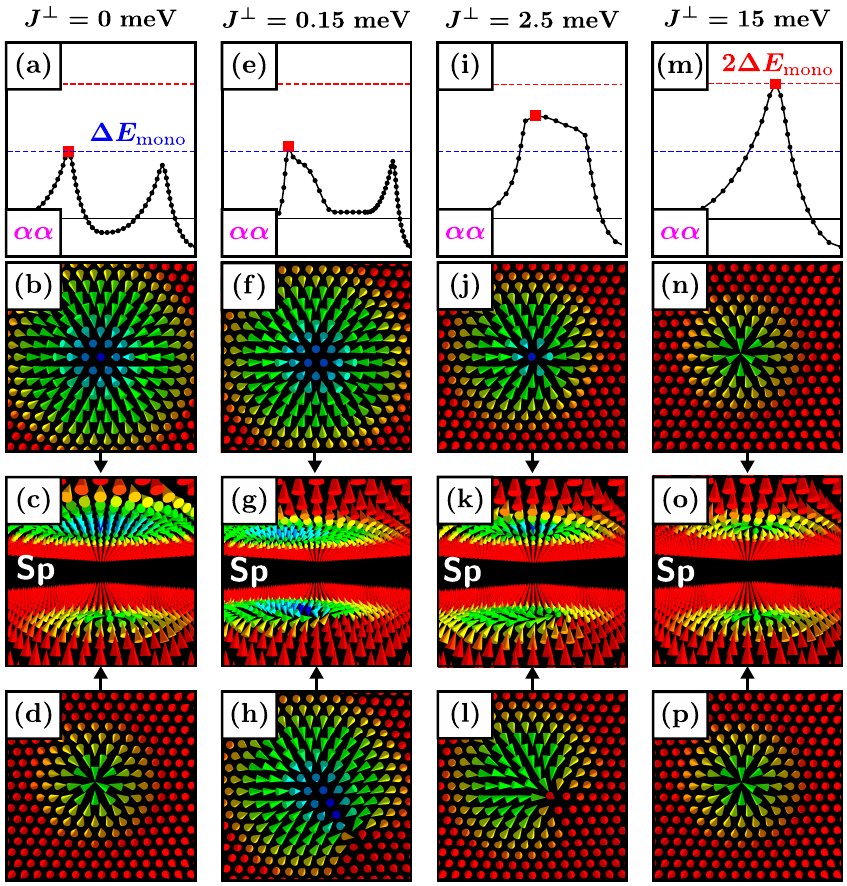}
\end{overpic}
\caption{Representation of collapse mechanisms for bilayer skyrmions in the $\alpha\alpha$-system for different interlayer exchange couplings $J^\perp$. The top row shows the total energy along the minimum energy path, while the first occurring saddle point is marked in red. The blue (red) dashed line represents the energy barrier (twice the energy barrier) of a skyrmion in the magnetic monolayer system Pd/Fe/Ir(111)\cite{malottki2017}. Below the spin configuration of the corresponding saddle point is visualized. (a-d) Successive radial collapses of the bilayer skyrmion for $J^\perp=0$ meV. (e-h) Successive chimera collapse for $J^\perp=0.15$ meV. (i-l) Chimera type collapse in one layer with shrunken skyrmion in the other layer for $J^\perp=2.5$ meV. (m-p) Simultaneous radial collapse for $J^\perp=15$ meV.}
\label{fig:overview_collapses}
\end{figure}

This section demonstrates how the interlayer exchange affects the collapse mechanisms of bilayer skyrmions in the $\alpha\alpha$- and $\alpha\beta$-system. We use the NRE~model throughout this section. Fig.~\ref{fig:overview_collapses} presents an overview over the variety of collapse mechanisms of bilayer skyrmions in the $\alpha\alpha$-system for different interlayer exchange couplings. The MEPs are shown in the top row with the spin configurations of the saddle point below. In the high interlayer exchange regime ($J^\perp=15$~meV, Fig.~\ref{fig:overview_collapses}(m-p)) we predict a bilayer skyrmion collapse with a single saddle point configuration which corresponds to twice the energy barrier of a skyrmion in the magnetic monolayer Pd/Fe/Ir(111)\cite{malottki2017}. The spin configuration obeys a radial collapse mechanism in both layers where three spins point towards each other. This collapse mechanism is widely investigated for magnetic monolayer skyrmions\cite{malottki2017,bessarab2018,muller2018,Heil2019} and agrees with the assumption in Eq.~(\ref{Eq:l_emono_conjecture}).

The other limit of the uncoupled system is displayed in Fig.~\ref{fig:overview_collapses}(a-d). Here the collapse of the bilayer skyrmion consists 
of two independent collapses of the skyrmions in the different layers each of them resembling the radial collapse of a skyrmion in the monolayer system. The energy barriers of both decays coincide with the energy barrier reported for the skyrmion in Pd/Fe/Ir(111)\cite{malottki2017}.

Collapse mechanisms for the intermediate coupling regime as displayed in Fig.~\ref{fig:overview_collapses}(e-h) for $J^\perp=0.15$~meV and in Fig.~\ref{fig:overview_collapses}(i-l) for $J^\perp=2.5$~meV already demonstrate the increased complexity as opposed to monolayer skyrmions. This regime yields saddle point configurations following the chimera collapse mechanism predicted 
recently \cite{Heil2019,Meyer2019,desplat2019}. During this collapse process the  radial  symmetric  magnetic  structure  of  the  skyrmion changes through tilting the spins in one part of the edge 
(Fig.~\ref{fig:overview_collapses}(h)). Meyer \textit{et al.} found meta-stable skyrmions at zero external magnetic field in the magnetic monolayer system Rh/Co/Ir(111) and predicted them to collapse via the chimera transition mechanism \cite{Meyer2019}. Very recently the chimera collapse of a skyrmion in a ultrathin magnetic film system was identified experimentally\cite{muckel2021}. Here, we observe the chimera transition as part of a successive decay of skyrmions in different layers for $J^\perp=0.15$~meV where the first transition presents a chimera saddle point configuration while the second skyrmion follows the radial mechanism.

For slightly higher interlayer exchange $J^\perp=2.5$~meV the MEP of the bilayer skyrmion collapse exhibits a single saddle point. The corresponding spin configuration shows a chimera-type configuration in one layer while a skyrmion of reduced size compared to the initial state is obtained for the other layer 
(Fig.~\ref{fig:overview_collapses}(j-l)).

In the following we analyze the MEPs for the bilayer skyrmions shown in Fig.~\ref{fig:overview_collapses} in detail to achieve understanding of the origin of the variety of the collapses. Although we discuss bilayer skyrmions in the $\alpha\alpha$-system these are representative for the corresponding skyrmions in the $\alpha\beta$-system as we find analog collapse mechanisms for the same interlayer exchange parameters there. Furthermore, it is worth mentioning that we
always find
two MEPs for collapse mechanisms which include changing first the magnetization in one layer followed by a change in the other layer as the order of the collapses is exchangeable.

\begin{figure}
\includegraphics[scale=1]{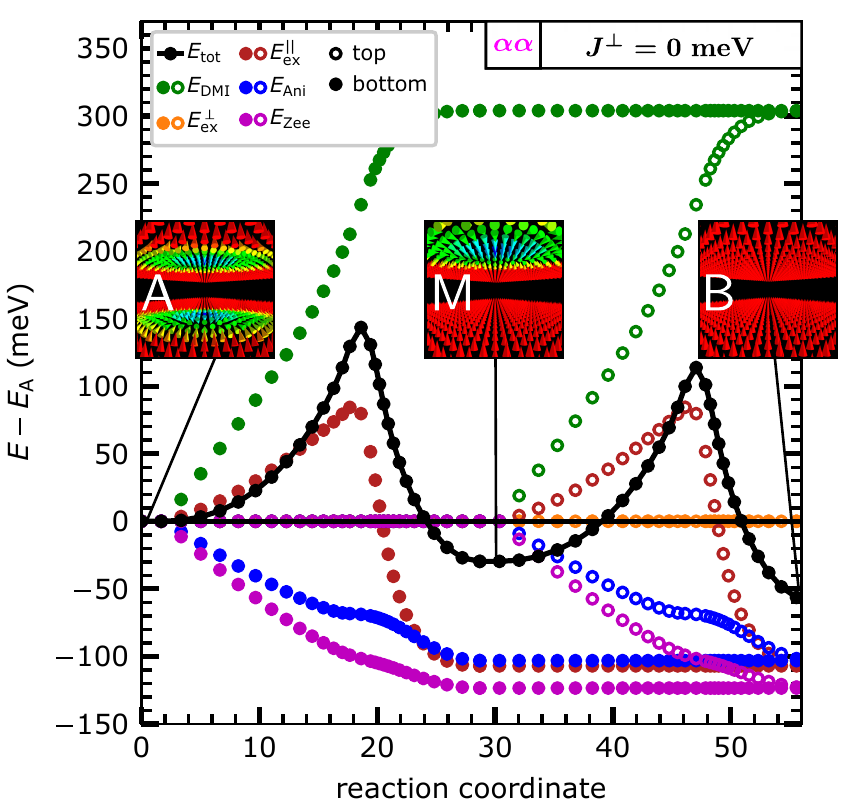}
\caption{Minimum energy path for a bilayer skyrmion in the $\alpha\alpha$-system without interlayer exchange coupling ($J^\perp=0.0$~meV). The total energy is plotted in black. The energy contributions of the different interactions are represented by the color code (see legend). Open circles represent the top Fe layer while filled circles symbolize the bottom Fe layer. The spin configurations of the initial bilayer skyrmion $A$, the intermediate minimum $M$ and the final field polarized state $B$ are shown in the insets.}
\label{fig:J_0_cubic_path}
\end{figure}

Starting with the uncoupled bilayer ($J^\perp=0$~meV) we 
decompose the total energy of the MEP
into the different energy contributions of Eq.~(\ref{gl:energy_model}) 
(Fig.~\ref{fig:J_0_cubic_path}). As highlighted in Sec.~\ref{ssec:method_GNEB} the MEP of a bilayer skyrmion in this system contains an intermediate minimum (M). This minimum is associated with a skyrmion in one Fe layer, which is unchanged concerning the corresponding layer for the A state, and one collinear aligned Fe layer. This indicates that the skyrmions in the different layers collapse independent from each other for the uncoupled Fe layers. Hence, we find two saddle point configurations with the energies $E_{\text{Sp}_1}$ and $E_{\text{Sp}_2}$, respectively. These energies correspond to energy barriers equal to the barrier of the skyrmion in the magnetic monolayer Pd/Fe/Ir(111) (Fig.~\ref{fig:overview_collapses}(a)).

Both the anisotropy and the Zeeman term favor a parallel out-of-plane alignment of the spins in both layers. 
For the sake of completeness we show all energy contributions to the total energy of the MEP in Fig.~\ref{fig:J_0_cubic_path}. In the following figures we restrict the decomposition to the intralayer exchange energy $E_\text{ex}^{\parallel}$ and the DMI energy $E_\text{DMI}$ since they dominate the energy of the saddle points $E_{\text{Sp}_{1/2}}$ and therefore the energy barriers (cf.~Fig.~\ref{fig:J_0_cubic_path}). Further the interlayer exchange energy $E_\text{ex}^\perp$ is included. 
Note, that the large contribution of $E_\text{ex}^{\parallel}$ to $E_{\text{Sp}_{1/2}}$ originates from the intralayer exchange frustration within the NRE model 
(Tab.~\ref{tab: Pd_Fe_Ir_parameters}), as reported in Ref.~\cite{malottki2017}. Below we will call this mechanism the successive radial (SR) collapse. We predict the SR collapse only for very low values of the interlayer exchange coupling.

\begin{figure}
\includegraphics[scale=1]{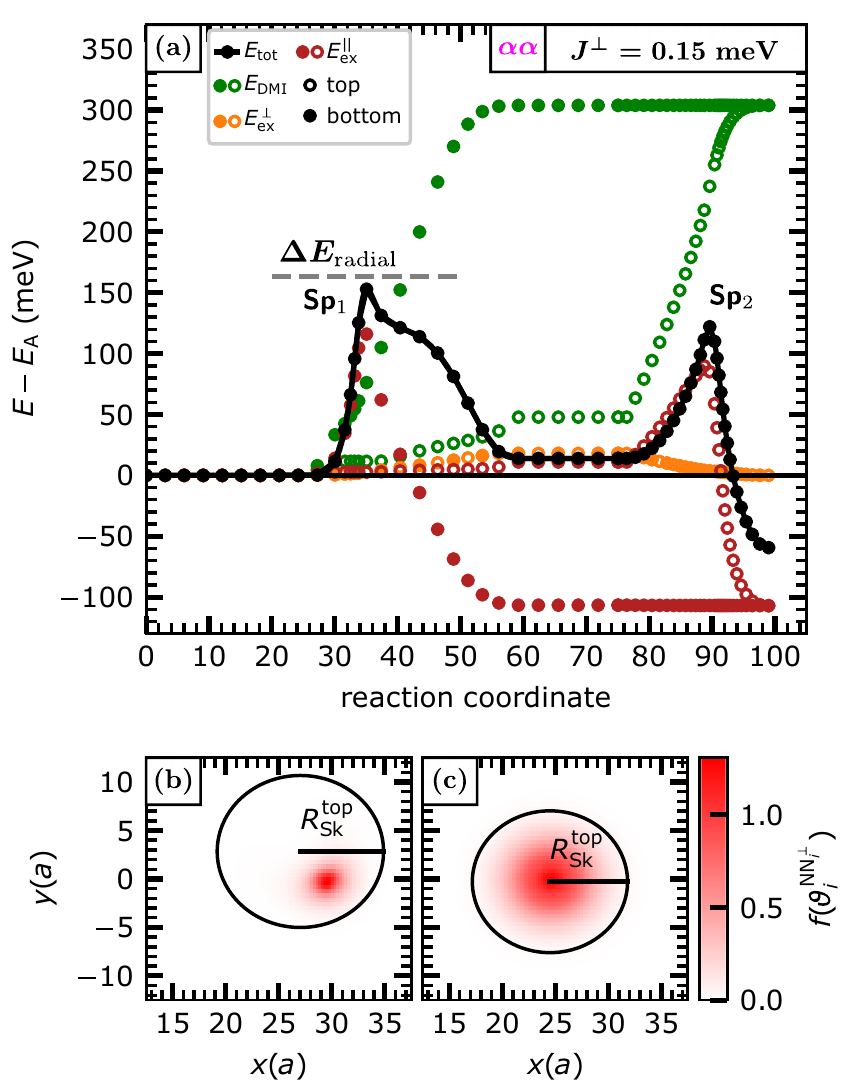}
\caption{(a) Minimum energy path for a bilayer skyrmion in the $\alpha\alpha$ system with interlayer exchange coupling ($J^\perp=0.15$~meV). The energy contributions of the different interactions are represented by the color code (see legend). Open circles represent the top Fe layer while filled circles symbolize the bottom Fe layer. The first saddle point corresponds to a chimera collapse 
(cf.~Fig.~\ref{fig:overview_collapses}(e-h)). The energy barrier corresponding to the meta-stable radial saddle point configuration is visualized 
by the dashed gray line. (b,c) The interlayer exchange energy costs $f(\vartheta_i^{\text{NN}_i^\perp})$ are presented by the color code 
versus the in-plane directions, where $a$ is the in-plane lattice constant. The skyrmion radius $R_\text{Sk}^\text{top}$ and the position of the unchanged layer (top layer in this example)
during the first part of the collapse is represented by a circle. 
While (b) belongs to the chimera collapse saddle point configuration (c) represents the saddle point of the meta-stable SR collapse for $J^\perp=0.15$ meV.}
\label{fig:J_0_15_cubic_path}
\end{figure}

Fig.~\ref{fig:J_0_15_cubic_path}(a) shows the MEP when one increases the interlayer exchange to $J^\perp = 0.15$~meV. Now the intermediate configuration M becomes less favorable due to the increasing interlayer energy costs. 
One can also recognize that the shape of the total energy of the MEP changes
for the first collapsing layer
with respect to that observed in Fig.~\ref{fig:J_0_cubic_path}.
This can be attributed to the appearance of the chimera collapse. The first part of the collapse (reaction coordinate $<25$) corresponds to a side wards movement of the initial skyrmion, which does not lead to an increase in the energy. This movement can be explained as a consequence of the initial geodesic path as described in Ref.~\cite{stephan_phd}. The energy barrier of the chimera collapse is dominated by the intralayer exchange, while the amount of the DMI energy at the saddle point is relatively low compared to the radial collapse mechanism. This is due
to the fact that the noncollinear alignment is preserved for the most part of the skyrmion and only the spins in one part of the margin of the skyrmion are tilted as visible in Fig.~\ref{fig:overview_collapses}(h). After the skyrmion in one layer has collapsed to a parallel alignment the second skyrmion follows a radial mechanism with the corresponding saddle point $\text{Sp}_2$.

Although the interlayer exchange energy does not contribute to the saddle point corresponding to the chimera collapse in one layer ($\text{Sp}_1$) it can explain the appearance of this collapse mechanism. If $N_{1}$ is the number of magnetic moments in layer $1$ the energy costs due to interlayer exchange can be written as:
\begin{align}
E_\text{costs}^\perp&=2\cdot J_1^\perp\sum\limits_{i=1}^{N_1}\sum\limits_{\text{NN}_i^\perp}\left(1-\mathbf{m}_i\cdot\mathbf{m}_{\text{NN}_i^\perp}\right)\notag \\
&=2\cdot J_1^\perp\sum\limits_{i=1}^{N_1}\sum\limits_{\text{NN}_i^\perp}\left(1-\cos\vartheta_i^{\text{NN}_i^\perp}\right)\notag \\
&=2\cdot J_1^\perp\sum\limits_{i=1}^{N_1}\sum\limits_{\text{NN}_i^\perp}f(\vartheta_i^{\text{NN}_i^\perp}),
\label{gl:interlayer_ex_costs}
\end{align}
while $i$ represents the magnetic moments of one layer, $\text{NN}_i^\perp$ numerates the next interlayer neighbors of the magnetic moment $i$. The angle between a magnetic moment $i$ and its neighbor $\text{NN}_i^\perp$ is expressed by $\vartheta_i^{\text{NN}_i^\perp}$ and the factor of two arises due to the definition of the exchange constant as per atom. The interlayer exchange costs are proportional to the introduced function $f(\vartheta_i^{\text{NN}_i^\perp})$. In this formulation it becomes visible that increased angles between the magnetic configurations of the different layers lead to increased interlayer exchange costs. Therefore, the intermediate minimum M becomes less favorable, when the interlayer exchange increases. 

 In Fig.~\ref{fig:J_0_15_cubic_path}(b) we visualize $f(\vartheta_i^{\text{NN}_i^\perp})$ for the saddle point configuration $\text{Sp}_1$ across the in-plane directions of the system, which is a direct measure for the interlayer exchange energy costs. These costs concentrate mainly on one point of the edge of the skyrmion where the spins are tilting as described above. The rest of the skyrmion is still parallel aligned to the nearly unchanged skyrmion in the other layer, which reduces the interlayer exchange costs. In Fig.~\ref{fig:J_0_15_cubic_path} the nearly unchanged skyrmion during the first part of the collapse corresponds to the top layer and is represented by its radius $R_\text{Sk}^\text{top}$. The radius was determined through applying the definition of Bogdanov and Hubert\cite{bogdanov1994} onto the skyrmion profile\cite{varentcova2018} gained through a fit to the magnetization of the top layer. 

The role of the interlayer exchange favoring the chimera saddle point can be underlined by a comparison with the SR collapse mechanism. For $J^\perp=0.15$~meV it is still possible 
within the simulation to meta-stabilize the SR collapse mechanism. As indicated by the dashed gray line in Fig.~\ref{fig:J_0_15_cubic_path}(a) the corresponding energy barrier of the SR collapse is slightly larger than the energy barrier of the chimera collapse. Fig.~\ref{fig:J_0_15_cubic_path}(c) shows the interlayer exchange costs of the SR collapse mechanism for $J^\perp=0.15$~meV and one can identify the increased energy costs due to the symmetric shrinking of the skyrmion in one layer compared to the asymmetric chimera collapse
(Fig.~\ref{fig:J_0_15_cubic_path}(b)). 
Comparing the radius of the skyrmion in the top layer ($R_\text{Sk}^{\text{top, chim}}=7.84a$) for the chimera collapse with the radius for the SR mechanism ($R_\text{Sk}^{\text{top, rad}}=7.32a$) the skyrmion in the top layer is slightly smaller for the SR collapse. Here $a$ is the in-plane lattice constant. This indicates that the radial collapse mechanism already involves a small part of simultaneous shrinking of both skyrmions in the first part of the collapse, which is also related to reducing interlayer exchange costs. 

It is noteworthy that the chimera collapse also occurs in the monolayer system but at lower magnetic fields\cite{stephan_phd}. Therefore the interlayer exchange interaction shifts the transition of the radial to the chimera collapse so that it can occur also at higher fields. In the following we assign the name successive chimera (SC) collapse to transitions which show a chimera collapse for the first layer followed by a radial collapse for the skyrmion in the other layer.

\begin{figure}
\includegraphics[scale=1]{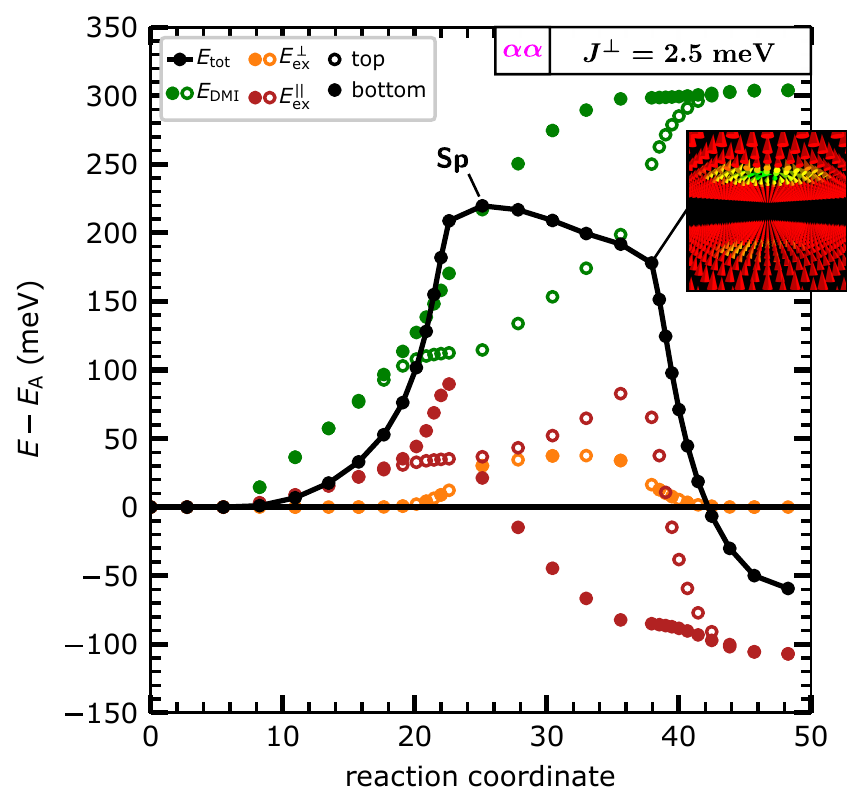}
\caption{Minimum energy path for a bilayer skyrmion in the $\alpha\alpha$-system with interlayer exchange coupling ($J^\perp=2.5$~meV). The energy contributions of the different interactions are represented by the color code (see legend). Open circles represent the top Fe layer while filled circles symbolize the bottom Fe layer. An exemplary spin configuration in the region of the collapse of the second skyrmion is shown as an inset. For the spin configuration of the actual saddle point see Fig.~\ref{fig:overview_collapses}(i-l)}.
\label{fig:J_2_5_cubic_path}
\end{figure}
Increasing the interlayer exchange to $J^\perp=2.5$~meV
(Fig.~\ref{fig:J_2_5_cubic_path}), we enter the regime of intermediate interlayer coupling. 
The initial GNEB calculations as described in Sec.~\ref{ssec:method_GNEB} 
do not show any intermediate minimum and 
the path has only one saddle point configuration (Sp).  This saddle point configuration includes a chimera saddle point
(cf.~Fig.~\ref{fig:overview_collapses}(l)) for one layer while the other layer has a radial structure of reduced radius compared to the initial configuration 
(cf.~Fig.~\ref{fig:overview_collapses}(j)). Thus the part of the collapse, which reduces the size of the skyrmion, 
occurs simultaneously in both Fe layers.
The region of the saddle point describes a successive chimera collapse of the skyrmion in one layer followed by a radial collapse of the skyrmion in the other layer. This is underlined by the inset in Fig.~\ref{fig:J_2_5_cubic_path}. Although this is not the saddle point configuration the second skyrmion collapse appears to be radial symmetric. To emphasize the fact that this collapse mechanism is partly simultaneous and partly successive we call this mechanism semi-successive chimera (SSC) collapse.

Comparing the interlayer exchange energy for the path for $J^\perp=2.5$~meV with the one for $J=0.15$~meV it is striking that it varies only slightly. The lifting of the intermediate minimum M occurs rather due to the more concurrent DMI~energy curves for the two layers. The difference between the 
DMI energy of the bottom and top layer along the reaction coordinate could thus be used as a quantity to define how simultaneous a collapse proceeds in the bilayer.

The SSC collapse mechanism changes to a different 
semi-successive mechanism for $J^\perp=4.9$~meV, where the transition of the bilayer skyrmion is simultaneous for most parts of the collapse but the region of the saddle point reveals two successive radial mechanisms. Due to the similarity of this transition to the SSC collapse we 
do not discuss this mechanism in detail here, but as it becomes important for the effective model later we assign the name semi-succesive radial (SSR) collapse.

\begin{figure*}
\includegraphics[scale=1]{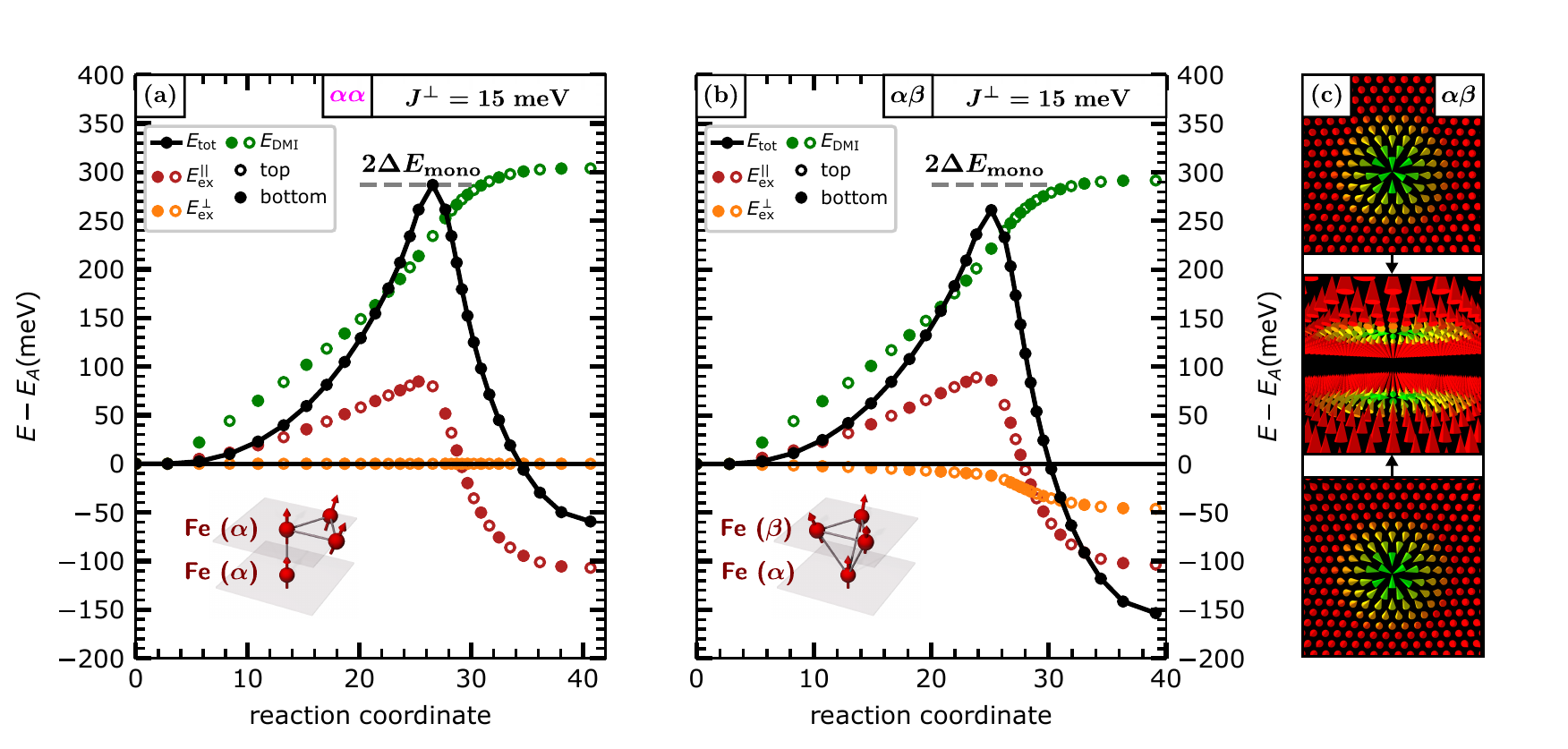}
\caption{(a) Minimum energy path for a bilayer skyrmion in the  $\alpha\alpha$-system with interlayer exchange coupling ($J^\perp=15.0$~meV). The energy contributions of the different interactions are represented by the color code (see legend). Open circles represent the top Fe layer while filled circles symbolize the bottom Fe layer. Twice the energy barrier of a skyrmion in the corresponding monolayer system Pd/Fe/Ir(111) is shown as dashed gray line. The inset displays schematically the stacking of the Fe layers. (b) Analog visualization of the minimum energy path for the $\alpha\beta$-system for $J^\perp=15$ meV. (c) The spin configuration of the saddle point for the collapse of the bilayer skyrmion in the $\alpha\beta$-system is shown. For the corresponding configuration of the $\alpha\alpha$-system see Fig.~\ref{fig:overview_collapses}(m-p).}
\label{fig:J_9_comp_J_7}
\end{figure*}
Finally increasing the interlayer exchange coupling to $J^\perp=15$~meV we end up in the high interlayer coupling regime for both the $\alpha\alpha$ and the $\alpha\beta$ system. The MEPs of the bilayer skyrmions in both systems are presented in Figs.~\ref{fig:J_9_comp_J_7}(a,b). In this regime, significant differences occur in the MEP between the $\alpha\alpha$ system and the $\alpha\beta$ system. We start with the description of the bilayer skyrmion in the $\alpha\alpha$-system in Fig.~\ref{fig:J_9_comp_J_7}(a). The difference between the DMI~energy of the bottom and top layers disappears, indicating a simultaneous collapse of both layers. The simultaneous change of both layers during the skyrmion collapse avoids interlayer exchange costs, which can be seen through the vanishing energy contribution $E_\text{ex}^{\perp}$. The consequence of this simultaneous collapse is that the energy barrier is equal to twice the energy barrier of a skyrmion in the monolayer system ($2\Delta E_\text{mono}$). Moreover, the mechanism in both layers corresponds to the radial collapse of the monolayer skyrmion.

If we compare this with the collapse of the bilayer skyrmion in the $\alpha\beta$ system (Fig.~\ref{fig:J_9_comp_J_7}(b)), we also find that the DMI energy contributions of both layers to the MEP are identical. Again, the collapse is simultaneous in both Fe layers. However, if we look at the energy barrier, we find a slight reduction compared to $2\Delta E_\text{mono}$, which is explained by the interlayer exchange. If we analyze the contribution of $E_\text{ex}^\perp$ to the MEP in Fig.~\ref{fig:J_9_comp_J_7}(b), we find that the saddle point is energetically favored over the initial state. Furthermore, the field-polarized state is clearly favored with respect to the interlayer exchange. The explanation for this is analogous to the cause of the shift of the critical fields in the magnetic phase diagram discussed in Sec.~\ref{ssec:phase_dia}. 

The insets in Fig.~\ref{fig:J_9_comp_J_7}(a) and (b) contrast the horizontal shift of the Fe layers in the case of the $\alpha\beta$ system with the directly superimposed layers of the $\alpha\alpha$ system. This shift causes noncollinear regions of magnetization within one layer to be slightly tilted with respect to the same structure in the other layer. Collinear regions are therefore favored in terms of interlayer exchange and in this sense the bilayer skyrmion is unfavorable relative to the field polarized state. Since the saddle point state has a smaller noncollinear fraction than the skyrmion, the energetic order with respect to interlayer exchange in the $\alpha\beta$-system results in $E_\text{ex}^\perp(\text{A})>E_\text{ex}^\perp(\text{Sp})>E_\text{ex}^\perp(\text{B})$. The collapse mechanism, on the other hand, is very similar for the bilayer skyrmions in the $\alpha\alpha$- (Fig.~\ref{fig:overview_collapses}~(m-p)) and $\alpha\beta$-system (Fig.~\ref{fig:J_9_comp_J_7}~(c)). Only the three central spins of the radial saddle point for the skyrmion in the $\alpha\beta$-system have a slightly larger out-of-plane fraction (See App.~\ref{sec:app_minmz}). We will call this collapse mechanisms for high interlayer exchange simultaneous collapse in the following.

\subsection{Energy barriers for bilayer skyrmions}
\label{ssec:barriers}
\begin{figure*}
\includegraphics[scale=1]{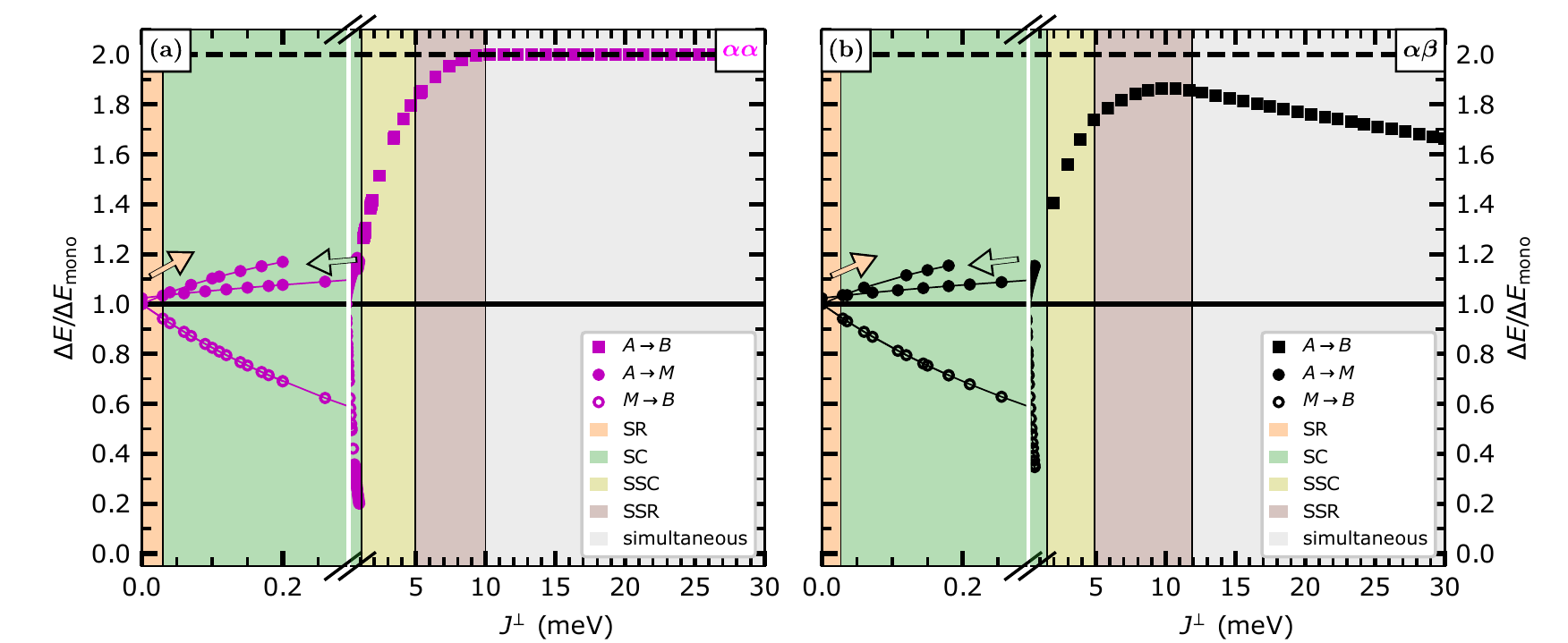}
\caption{(a) Energy barriers $\Delta E$ of bilayer skyrmions in the $\alpha\alpha$ system relative to the energy barrier of a skyrmion in the magnetic monolayer system Pd/Fe/Ir(111) for various interlayer exchange couplings $J^\perp$. For clarity the shown data point density is reduced and the $J^\perp$ axis has an enlarged scale for small values.
We used the NRE model for these calculations (cf.~Tab.~\ref{tab: Pd_Fe_Ir_parameters}). The background color represents the type of the collapse mechanism as described in App.~\ref{sec:app_minmz} and Fig.~\ref{fig:app_minmz}. For collapse mechanisms with two saddle point configurations the corresponding energy barriers of the first (second) collapse are labeled with filled (open) circles, while the energy barriers corresponding to a collapse mechanism with a single saddle point configuration are symbolized with squares. The solid (dashed) black line represents the energy barrier (twice the energy barrier) of a skyrmion in the magnetic monolayer system Pd/Fe/Ir(111). The arrows indicate the directions of the piecewise GNEB calculations as described in 
detail in the text. (b) Analog visualization to (a) for bilayer skyrmions in the $\alpha\beta$ system.}
\label{fig:compare_stacks_summarize}
\end{figure*}

To understand the role of interlayer exchange for
the stability of bilayer skyrmions, a detailed discussion of the corresponding energy barriers is inevitable (cf.~Eq.~(\ref{Eq:arrhenius})). We therefore systematically varied the interlayer exchange ($J^\perp\in[0, 30]$~meV) for bilayer skyrmions (A) in the $\alpha\alpha$- and $\alpha\beta$-systems and calculated the energy barriers for the collapse to the field-polarized state (B). As described in Sec.~\ref{ssec:introduction_to_mechanisms}, MEPs with an intermediate minimum occur in the low interlayer exchange coupling region. These MEPs are associated with two energy barriers. While the first barrier describes the transition of the skyrmion in one layer ($A\rightarrow M$), the second barrier is associated with the collapse of the skyrmion in the other layer ($M\rightarrow B$). In contrast, for high interlayer exchange, we find transitions of the bilayer skyrmion to the field polarized state of the bilayer with just one energy barrier ($A\rightarrow B$). Our goal is to study the energy barriers $\Delta E$ of bilayer skyrmions relative to the energy barrier of a skyrmion $\Delta E_\text{mono}$ in the magnetic monolayer system Pd/Fe/Ir(111)\cite{malottki2017}. Fig.~\ref{fig:compare_stacks_summarize} 
displays the ratio
$\Delta E/\Delta E_\text{mono}$ 
as a function of
$J^\perp$ for the $\alpha\alpha$ and $\alpha\beta$ system. To provide increased resolution for low $J^\perp$ in Fig.~\ref{fig:compare_stacks_summarize}, the corresponding axis was provided with two different scales. The collapse mechanisms introduced in Sec.~\ref{ssec:introduction_to_mechanisms} are illustrated by the background color in Fig.~\ref{fig:compare_stacks_summarize}. In the following, we will discuss the determination of these areas and the behavior of the energy barrier with increasing $J^\perp$.

For very low interlayer exchange, the SR collapse is preferred. This mechanism is associated with large interlayer exchange costs, as discussed in the context of Fig.~\ref{fig:J_0_15_cubic_path}. The SC collapse minimizes these costs and is therefore preferred for increasing interlayer exchange. However, it is possible to meta-stabilize the SR mechanism up to $J^\perp=0.2$~meV
as shown in Fig.~\ref{fig:compare_stacks_summarize}.
This was calculated using the following methodology. Since the GNEB method calculates the local MEP closest to the initial path, it is possible to increase (decrease) the interlayer exchange piecewise and always use the result of the previous GNEB calculation as the initial path for calculating the collapse for the next larger (lower) interlayer exchange. The 
orange arrows in Fig.~\ref{fig:compare_stacks_summarize} symbolize such calculations for the SR collapse starting from $J^\perp=0$~meV. The steps were chosen to be $\Delta J^\perp=0.01$~meV but for better visibility only a few data points are presented in Fig.~\ref{fig:compare_stacks_summarize}. Similarly, a calculation of the SC collapse starting from $J^\perp=0.3$~meV was performed for piecewise smaller interlayer exchange. This is indicated by the green arrows in Fig.~\ref{fig:compare_stacks_summarize}. From the intersection of the curve for the SR collapse and the curve for the SC collapse, the change of mechanism for $J^\perp=(0.03\pm 0.01)$~meV for the $\alpha\alpha$- and for $J^\perp=(0.027\pm 0.009)$~meV for the $\alpha\beta$ system is obtained, where the error results from the distance of the data points in the $J^\perp$ direction.

The further one increases the interlayer exchange, the more energetically unfavorable the intermediate minimum becomes. This leads to the fact that above a certain $J^\perp$ only MEPs with a single saddle point exist. This transition defines the change of the SC-collapse to the SSC mechanism. For the $\alpha\alpha$ system this happens at $J^\perp=(1.1\pm0.2)$~meV and for the $\alpha\beta$ system at $J^\perp=(1.5\pm0.6)$~meV, as indicated by the change of the background colors in Fig.~\ref{fig:J_9_comp_J_7}.

As discussed in Sec.~\ref{ssec:introduction_to_mechanisms}, a chimera-like saddle point is energetically favorable for successive collapsing skyrmions. Considering the SSC collapse mechanism for increasing interlayer exchange, we find that the magnetization changes in both layers become more and more similar during the collapse, except for the region of the saddle point (see Fig.~\ref{fig:J_2_5_cubic_path}). However, as the shrinkage of the skyrmion proceeds simultaneously
in both layers, the noncollinear part of the magnetization for the saddle point becomes smaller. Above a certain interlayer exchange, the saddle point size is small enough that the tilting of the spins at the edge discussed in the context of Fig.~\ref{fig:J_0_15_cubic_path}(b,c) for the chimera-like saddle point means only small savings of the interlayer exchange costs. From this point on, the SSR collapse is preferred. The corresponding limit of the regimes in Fig.~\ref{fig:compare_stacks_summarize} is indicated by renewed change of background color. However, the position of 
this transition cannot be inferred
from the behavior of the energy barrier, because the curve
in Fig.~\ref{fig:compare_stacks_summarize} is continuous. Instead, the central spins of the saddle point configurations are analyzed. This approach is described in App.~\ref{sec:app_minmz} and Fig.~\ref{fig:app_minmz}. For the $\alpha\alpha$ system as well as for the $\alpha\beta$ system the change of the regimes happens for $J^\perp=(4.9\pm 0.05)$~meV.

It is remarkable how closely the collapse mechanisms in the $\alpha\alpha$ and $\alpha\beta$ system match in the regimes discussed so far. Let us now consider the regime of SSR collapse. Here the energy barrier of the bilayer skyrmion reaches a maximum and the first differences between the $\alpha\alpha$-system and $\alpha\beta$-system appear. While the energy barrier for the bilayer skyrmions in the $\alpha\alpha$-system converges towards twice the value of the energy barrier of the skyrmion in the monolayer system, the curve for the $\alpha\beta$-system only reaches a maximum of about $\operatorname{max}(\Delta E_{\alpha\beta})\approx 1.86\Delta E_{\text{mono}}$ with a decrease afterwards. Increasing the interlayer exchange further finally leads to the simultaneous collapse regime. The determination of the border is again described in App.~\ref{sec:app_minmz} and we observe the change for $J^\perp=(10.0\pm 0.05)$~meV for the $\alpha\alpha$- and for $J^\perp=(11.9\pm 0.05)$~meV for the $\alpha\beta$-system. 

The decrease in the energy barrier for skyrmions in the $\alpha\beta$ system occurs already before the transition to the completely simultaneous collapse mechanism happens. As the interlayer exchange is increased
within the simultaneous regime for the $\alpha\beta$ system the difference between the saddle point configuration and the bilayer skyrmion in terms of interlayer exchange energy increases favoring the saddle point. This leads to a linear decline of the energy barrier as the spin configurations along the MEP
do not change anymore in this regime but only the interlayer exchange constant $J^\perp$ varies linear the energy in Eq.~(\ref{gl:energy_model}). This is in sharp contrast to the behavior of the bilayer skyrmions in the $\alpha\alpha$ system. Here, the interlayer exchange energy contribution to the MEP reduces to zero when the collapse is simultaneous in both layers as all neighbors coupled via interlayer exchange are aligned parallel. Therefore, the energy barrier of the bilayer skyrmion equals twice the monolayer skyrmion energy barrier and is not affected by further changes in $J^\perp$.

From the decrease of the energy barrier of the skyrmion in the $\alpha\beta$-system for high interlayer exchange couplings we can draw the conclusion that stability of bilayer skyrmions not inevitable enlarges for increased interlayer exchange. Based on these results, it is important to understand for which interlayer exchange coupling $J_C^\perp$ a fully simultaneous collapse of the bilayer skyrmion occurs. The detailed investigation of these critical interlayer exchange parameters is given in Sec.~\ref{ssec:stacks_effective_model}.

\subsection{Critical interlayer exchange couplings}\label{ssec:stacks_effective_model}
\begin{figure*}
\includegraphics[scale=1]{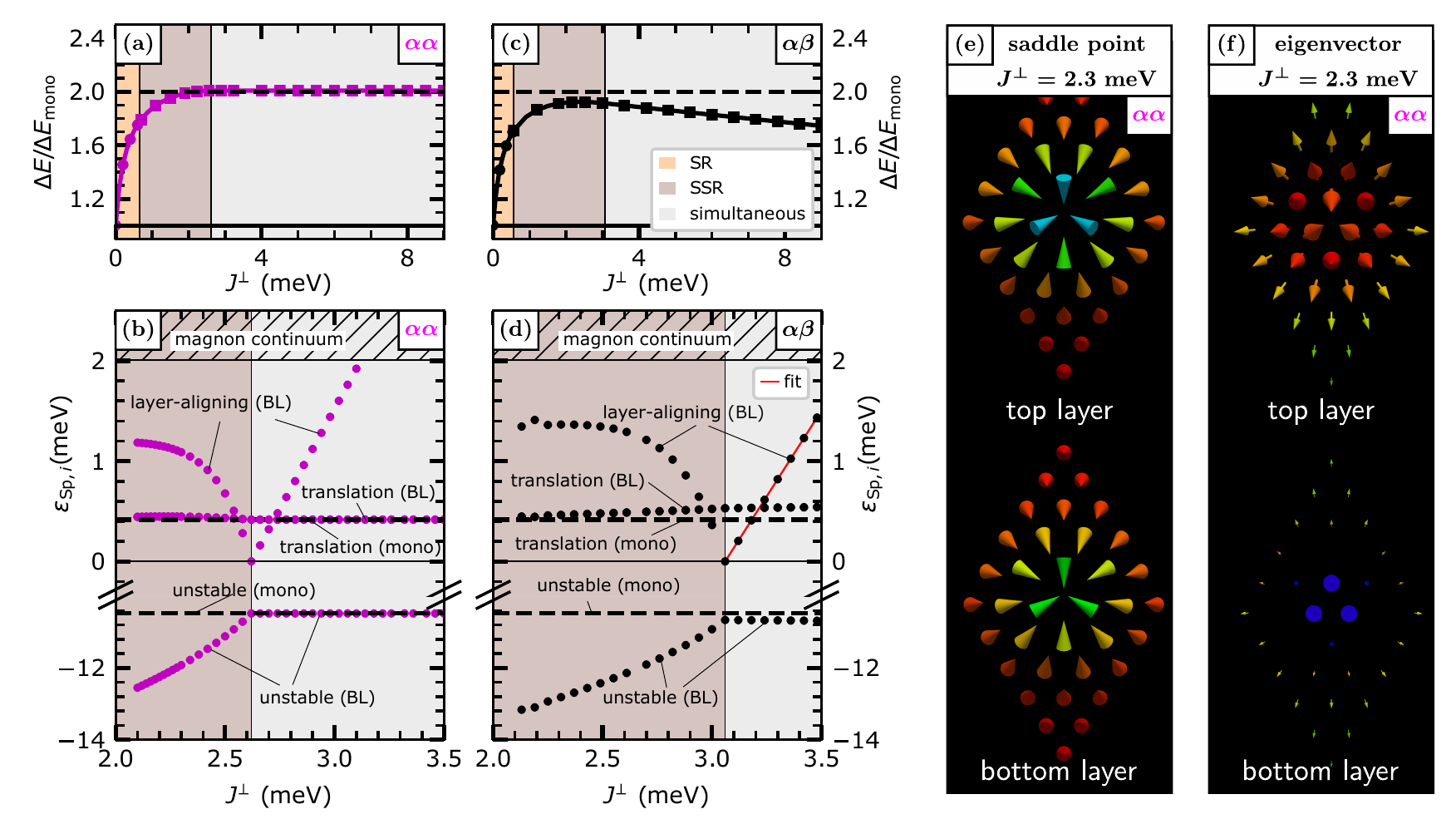}
\caption{(a,c) Energy barriers of bilayer skyrmions in the $\alpha\alpha$ and the $\alpha\beta$ system  relative to the energy barrier of skyrmions within the magnetic monolayer system Pd/Fe/Ir(111) for various interlayer exchange constants $J^\perp$. The effective parameter set (cf.~Tab.~\ref{tab: Pd_Fe_Ir_parameters}) is used and the background colors follow the definition of the collapse mechanisms as defined in Fig.~\ref{fig:compare_stacks_summarize} for the NRE model. For better visibility every third data point is displayed. (b,d) Visualization for the $\alpha\alpha$ and 
$\alpha\beta$ system, respectively, of the lowest eigenvalues of the Hessian of the saddle point configuration which belong to the energy barriers displayed in (a,b) for various $J^\perp$. 
The identified eigenmodes calculated for the bilayer system are labeled with BL. For a detailed description see the 
text. The dashed lines indicate the eigenvalues for a skyrmion in the monolayer system at $B=4.0$~T described with the effective model\cite{malottki2019}. In (d) a fit following Eq.~(\ref{gl:landau_curvature}) is presented by a red line. (e) Representation of the saddle point configuration for the bilayer skyrmion in the $\alpha\alpha$ system for $J^\perp=2.3$~meV. (f) Visualization of the eigenvector for the layer-aligning mode of the saddle point presented in (e). The color code represents the z-component of the orientation of the vectors.}
\label{fig:eig_transition}
\end{figure*}

During the preceding section the question arises for which interlayer exchange $J^\perp$ the collapse of bilayer skyrmions becomes fully simultaneous and which underlying physical properties determine this transition. %
To answer these questions, we reduced the complexity of the system by turning to the more simple representation of the intralayer interaction in effective nearest-neighbor approximation, with a value of $J_{1}^{||}=3.68$~meV, as reported in Ref.~\cite{malottki2017}. %
This excludes the effect of exchange frustration on the energy barrier which is now solely dominated by the DMI, with $D_\text{1}=1.39$~meV (See Tab.~\ref{tab: Pd_Fe_Ir_parameters}). With these parameters, we performed calculations of the magnetic bilayer system analog to the preceding section, yielding the energy barriers, $\Delta E$, over varying interlayer exchange coupling, $J^\perp$, for both the $\alpha\alpha$ and the $\alpha\beta$ stacking as displayed in Fig.~\ref{fig:eig_transition} (a) and (c), respectively.  %
Similar to the case of frustrated intralayer exchange interaction, we observe an initially strong increase and a subsequent convergence of the energy barrier to twice the value of the corresponding monolayer system for the $\alpha\alpha$-stacked bilayer. %
This value is again not reached by the skyrmion annihilation in the $\alpha\beta$ stacking, as the barrier starts to decrease with $J^{\perp}$ after a maximum has been reached around $J^\perp\approx2$~meV. %

Note, that within nearest-neighbor approximation no chimera collapse mechanism occurs in the low and intermediate interlayer exchange regimes, highlighting the crucial role of the intralayer exchange frustration for the formation of the chimera saddle point state \cite{Meyer2019,muckel2021,stephan_phd}. %
Without this additional stabilization, the energy difference between the radial symmetric and chimera saddle point structures in the monolayer system is larger than the potential energy gain of an occurring chimera saddle point in the bilayer skyrmion collapse. 
This demonstrates that frustration effects of the intralayer interactions can increase the complexity and variety of transitions in magnetic bilayer systems.

In the following we focus on the eigenspectra of saddle point states in the interlayer exchange interval $J^\perp\in[2.0, 3.5]$~meV, in which the transition of the semi-successive radial collapse (SSR) to the completely simultaneous radial collapse takes place. %
The eigenvalues of the Hessian $\mathcal{H}_\text{Sp}$ correspond to the curvature of the energy landscape in the vicinity of the saddle point in the basis of the eigenvectors.
In Fig.~\ref{fig:eig_transition}~(b), the spectra of the eigenvalues, $\epsilon_{\text{Sp},i}\in\{\epsilon_{\text{Sp},1},\dots,\epsilon_{\text{Sp},N}\}$, 
are shown for the saddle points of the $\alpha\alpha$-stacked bilayer versus
$J^\perp$. %
The eigenvalues of the monolayer system are added as a reference and agree with the eigenvalues published in Ref.~\cite{malottki2019, stephan_phd}. %

Both transition mechanisms exhibit a first order saddle point as they have exactly one negative eigenvalue shown in the lower part of the panel. %
The negative eigenvalue of the SSR mechanism increases with $J^\perp$ until it reaches the value of the monolayer close to the critical interlayer exchange of $J_C^\perp\approx2.6$~meV. %
In the regime of simultaneous coupling, the eigenvalue of the unstable mode lies exactly on the value of the monolayer, which can be expected since the magnetic structures of both layers are identical with the monolayer saddle point structure.%

In comparison to the monolayer system, a new saddle point eigenmode appears in the bilayer system, which connects the SSR and the simultaneous collapse mechanisms and is therefore coined layer-aligning mode (Fig.~\ref{fig:eig_transition}(b)).
For increasing $J^\perp$, its eigenvalue approaches zero at $J_C^\perp$ before it steeply rises again in the simultaneous collapse regime. %
This mode softening around $J_C^\perp$ is responsible for the transition between the SSR and the simultaneous collapse mechanisms in both the $\alpha\alpha$ and the $\alpha\beta$-stacking. %
The spectrum of the latter is shown in Fig.~\ref{fig:eig_transition} (d). %
It resembles the spectrum of the $\alpha\alpha$-stacking except for a larger critical interlayer exchange of $J_C^\perp\approx3.0$~meV and eigenvalues that slightly deviate from their monolayer counterparts with increasing $J^\perp$. %

In order to deepen the understanding of the layer-aligning mode, we display the spin structure of the SSR saddle point 
for a value of $J^\perp=2.3$~meV in Fig.~\ref{fig:eig_transition}~(e).
The spin structure of both layers is quite similar, but shows small deviations especially in the three central spins, which are slightly rotated downward in the top layer, but point almost toward each other in the bottom layer, implying that radial collapse is more advanced in the bottom layer than in the top layer as it is expected for the SSR collapse mechanism. %

By looking at the corresponding
eigenvector (Fig.~\ref{fig:eig_transition}(f)), one can already guess that 
its application to the top layer would push the magnetic structure in this layer further in the direction of the radial collapse. %
In contrast, the application of the eigenvector to the bottom layer would rotate the central moments in the opposite direction, resulting in more similar saddle points and thus a more simultaneous collapse in both layers. %

However, the visual examination of the eigenvector is limited and we apply the mode following method as proposed in Ref. \cite{stephan_phd}. %
Each mode following step consists of the calculation of the desired eigenvector by partial diagonalisation of the Hessian matrix and the subsequent rotation of the magnetic structure in the direction of this eigenvector. %
The resulting magnetic state is then the starting point for the next mode following step. %
A mode tracking algorithm which compares the previous eigenvector with the newly calculated ones ensures that always the eigenvector that is the most similar to the followed eigenmode is chosen. %
With this technique, the energy landscape in the direction of the eigenmode can be determined. %
See movies in the Supplemental Material for a visualization of this technique\cite{*[{See Supplemental Material }] [{for movies visualizing the application of mode following technique for the layer-aligning mode.}] SuppMov}.

\begin{figure}
\includegraphics[scale=1]{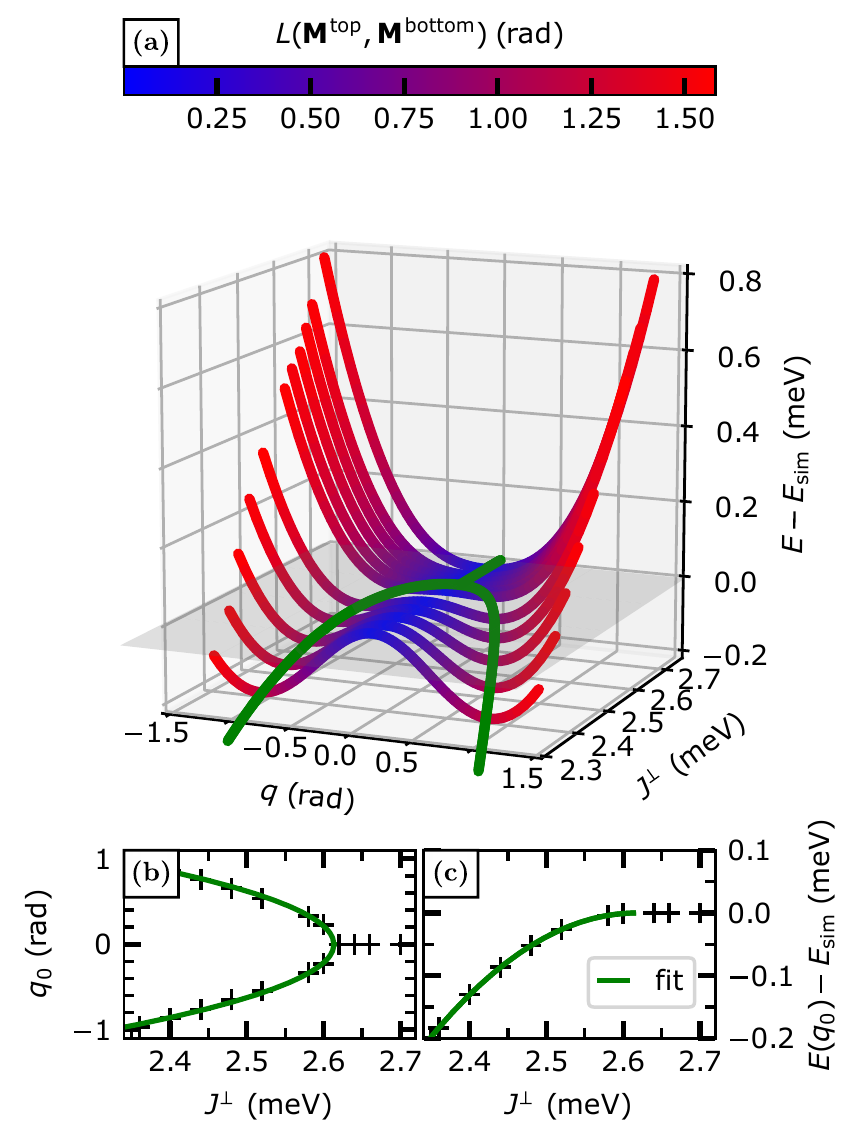}
\caption{(a) Local energy landscape along the layer-aligning mode for different $J^\perp$ around the saddle point configurations of bilayer skyrmions in the $\alpha\alpha$-system. An example for a corresponding eigenvector is shown in Fig.~\ref{fig:eig_transition}(f). The energy is displayed relative to the energy of the simultaneous saddle point configuration $E_\text{sim}$ and visualized over the displacement $q$ along the mode. The color code represents the geodesic distance between the magnetizations of the top layer $\vec{M}^\text{top}$ and the the bottom layer $\vec{M}^\text{bottom}$. (b,c) Position and value of the local energy minima from (a) recorded over $J^\perp$. The purple lines indicate fits of Eq.~(\ref{gl:landau_minpos}) and Eq.~(\ref{gl:landau_minenergy}), respectively.}
\label{fig:transitionmode}
\end{figure}

Fig.~\ref{fig:transitionmode} (a) shows the energy over the coordinate $q$, which determines the displacement of the magnetic structure along the layer-aligning mode, where a value of $q=0$ corresponds to the simultaneous collapse. %
The color encodes the geodesic distance between the magnetic structures in the top and bottom layer. %
Thus, the more blue (red) the color is, the more simultaneous (successive) the collapse mechanism is. The mode following calculations are performed for varying values of the interlayer exchange, $J^\perp$, resulting in one line per calculation. %
As starting points, the relaxed saddle point structures as obtained by CI-GNEB have been used. %

For small values of $J^\perp$, the energy profiles show two degenerate minima for both possible realizations of the SSR collapse mechanism. %
By following the layer-aligning mode from one minimum to the other, the saddle point of the simultaneous collapse is passed as an intermediate local energy maximum. %
With increasing $J^\perp$, the two degenerate energy minima become more shallow until they vanish at $J_C^\perp\approx2.6$~meV and a single minimum at $q=0$~rad emerges for even larger $J^\perp$.%

This behavior can be discussed analog to Landau's Theory for continuous phase transitions by modeling the energy to the fourth power along the mode:
\begin{align}
    E(q, J^\perp)-E_\text{sim} = a(J^\perp)\cdot q^2+\frac{b(J^\perp)}{2}q^4,
    \label{gl:landau_energy}
\end{align}
where the displacement along the mode $q$ takes the role of the ordering parameter, $E(q, J^\perp)$ is the energy along this ordering parameter for some value $J^\perp$ of the parameter provoking the phase transition and $E_\text{sim}$ is the zero point of this energy, which will be defined below.
In order to prohibit indefinite negative energies for indefinite order parameters $b(J^\perp)> 0$ has to hold and it will be further assumed that $b(J^\perp)=b_0$ is valid near $J^\perp$. Calculating the stationary points $q_0$ of Eq.~(\ref{gl:landau_energy}) yields:
\begin{align*}
    q_0^2 = -\frac{a}{b_0}.
\end{align*}
We obtain one local minimum ($q_0=0$) for $a>0$ and two local minima for $a<0$, which mimics exactly the behaviour of the energy landscape of the layer-aligning mode near $J_C^\perp$. Therefore, one can model $a(J^\perp)\approx a_0(J^\perp-J_C^\perp)$ for $a_0>0$ and $J^\perp$ close to $J_C^\perp$ and the positions of the minima follow
\begin{align}
    q_{0,\pm}=\pm \frac{a_0}{b_0}|J^\perp-J_C^\perp|^{\frac{1}{2}}.
    \label{gl:landau_minpos}
\end{align}
Further, the energy of the local minima can be determined through
\begin{align}
    E(q_{0,\pm})=-\frac{a_0^2}{2b_0}(J^\perp-J_C^\perp)^2.
    \label{gl:landau_minenergy}
\end{align}
Fitting Eq.~(\ref{gl:landau_minpos}) and Eq.~(\ref{gl:landau_minenergy}) to the data obtained by the mode-following method yields $a_0=(2.98\pm0.04)~\text{meV}/\text{rad}^2$, $b_0=(1.61\pm0.03)~\text{meV}/\text{rad}^4$ and $J_C^\perp=(2.613\pm0.003)~\text{meV}$ (See Fig.~\ref{fig:transitionmode}~(b),(c)). A phase transition implies a symmetric configuration above $J_C^\perp$ which splits up into two configurations with lower symmetry below $J_C^\perp$. The nature of this symmetry can be revealed through visualizing the geodesic distance\cite{bessarab2015} between the magnetization of the top Fe layer $\vec{M}^{\text{top}}$ to the magnetization of the bottom Fe layer $\vec{M}^{\text{bot}}$:
\begin{align}
L(\vec{M}^\text{top},\vec{M}^\text{bot})=\sqrt{(l_1^{\text{top,bot}})^2+(l_2^{\text{top,bot}})^2+\dots+(l_{N/2}^{\text{top,bot}})^2},
\label{eq:linfit}
\end{align}
where $N/2$ is the number of spins per layer and the $l_i^{\text{top, bot}}$ are geodesic distances between the points of the unit sphere, which correspond to the spins in the top and bottom layer, respectively. This quantity is represented by the color code in Fig.~\ref{fig:transitionmode}(a). While blue represents parallel aligned layers, red indicates a net angle between the magnetization of the different layers. Therefore one can conclude that indeed the simultaneous collapse mechanism matches with the high symmetry configuration for interlayer exchange couplings above $J_C^\perp$. Below $J_C^\perp$ two collapse mechanisms are possible with saddle point configurations obeying a successive transgression of the Bloch-like points in each layer and thus representing a lower symmetry. 
Note, that the energy for each slice (each $J^\perp$) in Fig.~\ref{fig:transitionmode}(a) is meant relative to the simultaneous configuration $E_\text{sim}$. This simultaneous configuration is a local minimum for $J^\perp>J_C^\perp$ and a local maximum for $J^\perp < J_C^\perp$. The displacement along the mode $q$ is also expressed relative to this simultaneous configuration.
All these consideration were done for the bilayer skyrmion collapse within the $\alpha\alpha$-system. For the purpose of substantiating the same mechanism in the $\alpha\beta$-system, we show that $a_0$ and $J^\perp$ can already be derived from the eigenvalue spectrum in Fig.~\ref{fig:eig_transition}(c). The second derivative of Eq.~(\ref{gl:landau_energy}) yields the curvature at the minimum along the energy reach along the layer-aligning mode $c$ and thus the corresponding eigenvalue 

\begin{align}
    \epsilon_{\text{Sp},c} = \begin{cases}2a_0|J^\perp -J_C^\perp|, &J^\perp >J_C^\perp\\ -4a_0|J^\perp -J_C^\perp|, &J^\perp<J_C^\perp\end{cases}
    \label{gl:landau_curvature}
\end{align}
for $J^\perp$ close to $J_C^\perp$. A fit of Eq.~(\ref{gl:landau_curvature}) to the layer-aligning mode for $J^\perp>J_C^\perp$ results $a_0=1.71~\text{meV}/\text{rad}^2$ and $J_C^\perp=3.06$~meV for the $\alpha\beta$ system. This fit is displayed by a line
in Fig.~\ref{fig:eig_transition}~(d).

\subsection{Varying the monolayer barrier}
\label{ssec:vary_monolayer_barrier}
\begin{figure}
\includegraphics[scale=1]{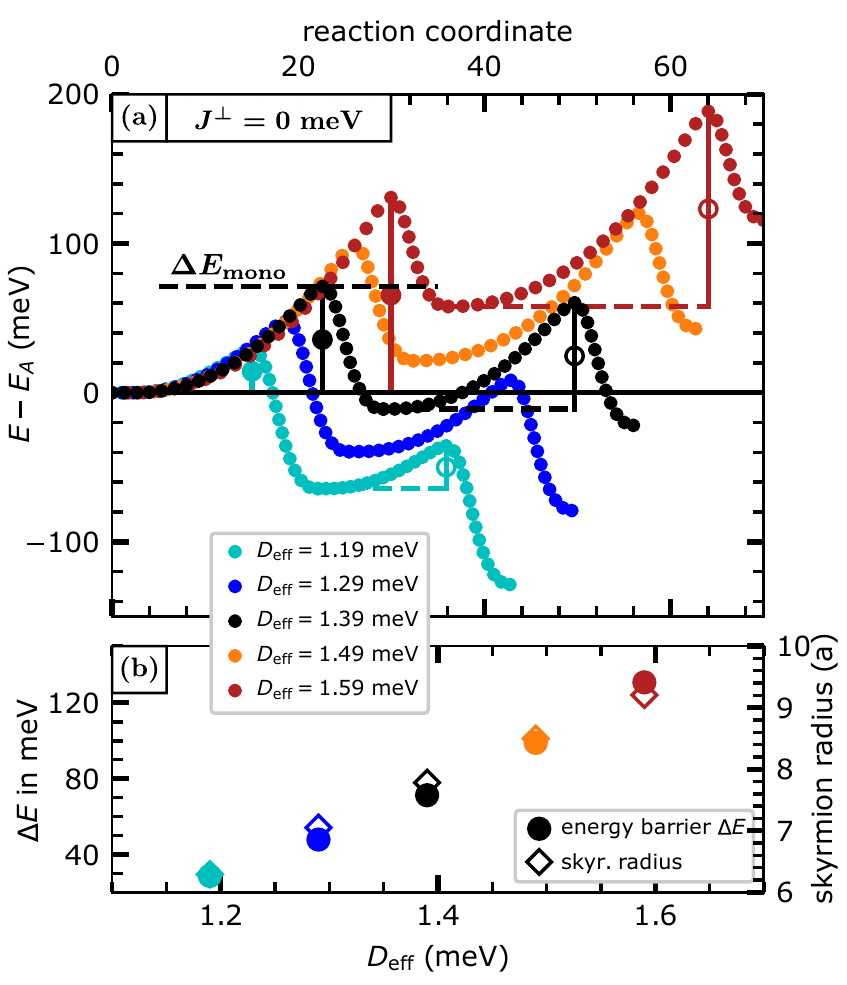}
\caption{(a) Minimum energy paths of bilayer skyrmions in the $\alpha\alpha$-system based on the effective parameters from the monolayer system Pd/Fe/Ir(111) for different values of the DMI (see color code in legend). The interalyer exchange constant is set to zero ($J^\perp=0.0$~meV). For comparison the energy barrier of the skyrmion in the magnetic monolayer for $D_\text{eff}=1.39$~meV\cite{malottki2017} is indicated as dashed line. (b) Energy barriers and radius of the bilayer skyrmions in the magnetic monolayer systems for different values of the DMI. The radius is given in units of the lattice constant $a$.}
\label{fig:zero_paths}
\end{figure}

Addressing the issue of designing a magnetic bilayer system which yields maximum skyrmion stability an estimation of 
the critical interlayer exchange strength
$J_C^\perp$ from the properties of the underlying monolayer system is important. We assume that the energy barrier of the skyrmion in the magnetic monolayer system may influence 
$J_C^\perp$ of the bilayer system. Therefore, we 
varied the barrier of each skyrmion in the bilayer 
by systematically varying
the DMI within the effective model ($D_\text{eff}\in[1.19, 1.59]$~meV).
Fig.~\ref{fig:zero_paths}(a) shows the obtained MEPs for five values of the DMI for $J^\perp=0$~meV. The collapses are similar to the MEP of the skyrmion in the underlying monolayer system as discussed in Fig.~\ref{fig:J_0_cubic_path}.
Fig.~\ref{fig:zero_paths}(b) presents the energy barrier of the first collapse and the radius of bilayer skyrmions for switched off interlayer exchange depending on the corresponding value of the DMI. In agreement with Ref.~\cite{varentcova2018} the radius and the energy barrier increase as the DMI strengthens. Since $J^\perp=0$~meV this energy barrier corresponds to the energy barrier of the underlying magnetic monolayer system $\Delta E_\text{mono}$. Therefore, the variation of the DMI-parameter yields a variation of the energy barrier of the magnetic monolayer skyrmion in the interval $\Delta E_\text{mono}\in[25, 130]$~meV.

\begin{figure}
\includegraphics[scale=1]{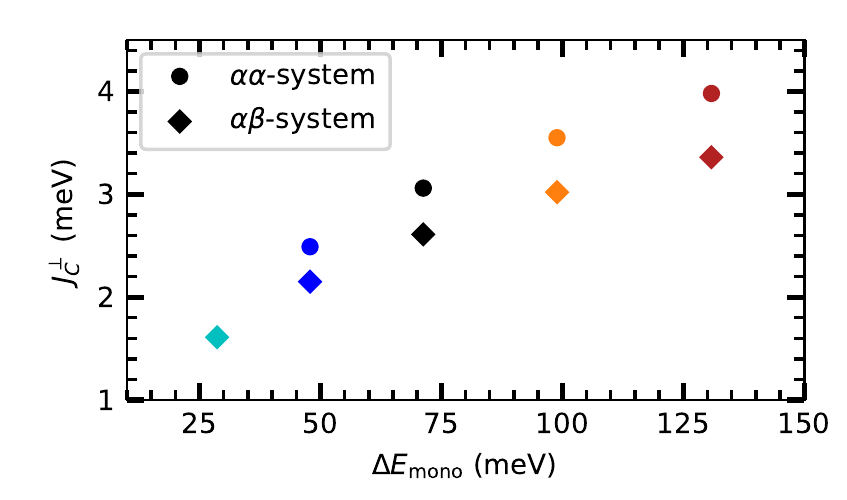}
\caption{Critical interlayer exchange $J_C^\perp$ for skyrmions in magnetic bilayer systems based on the effective parameter set (Tab.~\ref{tab: Pd_Fe_Ir_parameters}) for different values of the DMI. The color code indicates the different values of the DMI which define the energy barrier of a skyrmion in the corresponding magnetic monolayer system 
(cf.~Fig.~\ref{fig:zero_paths}). 
Circles denote the $\alpha\alpha$ system and diamonds represent the values of $J_C^\perp$ 
for the $\alpha\beta$ system. See Fig.~\ref{fig:app_vary_mono} for the corresponding visualization of the energy barriers of the bilayer skyrmions.}
\label{fig:tune_barrier}
\end{figure}

After defining $\alpha\alpha$- and $\alpha\beta$-stacked systems for these DMI values, we vary the interlayer exchange coupling and calculate the energy barriers of the bilayer skyrmions analog to Sec.~\ref{ssec:barriers}. We have to mention that during this variation no chimera type saddle points appear, which we attribute to the lack of intralayer frustration for the effective parameter set. In App.~\ref{sec:app_vary_mono} in Fig.~\ref{fig:app_vary_mono} we present, similar to Fig.~\ref{fig:eig_transition}(a) and (c), the energy barriers of the bilayer skyrmions relative to the energy barrier of the skyrmion in the corresponding monolayer system for the $\alpha\alpha$- and $\alpha\beta$-stacks, respectively.
Further we determined the critical interlayer exchange parameters $J_C^\perp$ by calculating the eigenvalue spectrum and applying a fit following Eq.~(\ref{gl:landau_curvature}) for $J^\perp>J_C^\perp$ as presented in Fig.~\ref{fig:eig_transition}(c). The obtained values of 
$J_C^\perp$ are displayed for the $\alpha\alpha$ and $\alpha\beta$ system in Fig.~\ref{fig:tune_barrier} as a function
of the energy barrier of a skyrmion in the corresponding monolayer system $\Delta E_\text{mono}$. 
Note, that the determination of $J_C^\perp$ for $D_\text{eff}=1.19$~meV was not possible for the $\alpha\beta$ system as the divergence of the layer-aligning mode is overlapping in the eigenvalue spectrum with another collapse mechanism of the low interlayer exchange regime here.

Although the critical parameter $J_C^{\perp}$ is always a bit larger for skyrmions in the $\alpha\beta$-stacked system than for the skyrmions in the $\alpha\alpha$-system 
both follow the same trend. As the monolayer barrier increases a higher interlayer exchange coupling is needed to force the system into a simultaneous collapse, which is indicated by the increase of $J_C^\perp$ in Fig.~\ref{fig:tune_barrier}. For comparison we observed $J_C^\perp=10.0$~meV for the $\alpha\alpha$ and $J_C^\perp=11.9$~meV for the $\alpha\beta$ system treated with the NRE-parameter set 
in Sec.~\ref{ssec:barriers}. This corresponded to an energy barrier $\Delta E_\text{mono}\approx 143$~meV of the underlying monolayer system. 

It is striking that for systems examined with the effective parameter set  the critical interlayer exchange parameters are significantly smaller than for the systems treated with the NRE model. This is an indication that in real systems with exchange frustration a much larger interlayer exchange is needed to force a simultaneous collapse of the skyrmions in the different layers. Therefore if one aims to design a magnetic bilayer system with maximum skyrmion stability two aspects have to be considered. On the one hand a higher energy barrier of a skyrmion in the underlying monolayer system provides a higher energy barrier for the simultaneously collapsing bilayer skyrmion. On the other hand one needs higher interlayer couplings to realize this simultaneous transition. 

\subsection{Energy barriers for multilayer skyrmions}
\label{ssec:beyond2lay}
\begin{figure}
\includegraphics[scale=1]{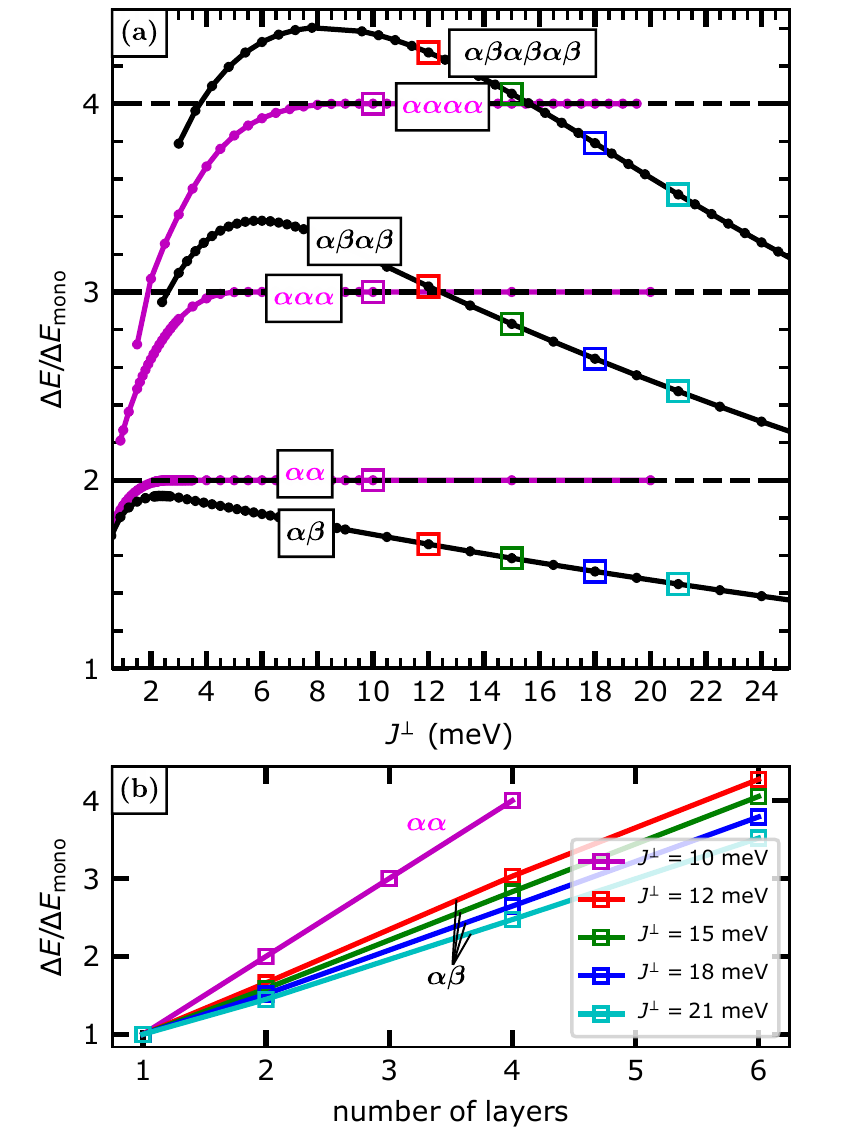}
    \caption{(a) Energy barriers of skyrmions in magnetic multilayer systems for various interlayer exchange couplings $J^\perp$ relative to the energy barrier of the skyrmion in the magnetic monolayer Pd/Fe/Ir(111). The effective parameter set is used for these calculations. The energy barriers for the skyrmions in the two-, three- and four-layer systems obeying a $\alpha\alpha$-stacking sequence are shown in magenta. The barriers for the two-, four- and sixlayer systems with $\alpha\beta$-stacking are shown in black. (b) Energy barriers from (a) as a function of the number of layers for fixed values of $J^\perp$. The corresponding data points are indicated by empty squares in (a). For the $\alpha\alpha$ systems the layer-dependent energy barrier is presented for $J^\perp=10$~meV, 
    while it is shown for
    the $\alpha\beta$ systems 
    for $J^\perp=12$~meV, $J^\perp=15$~meV, $J^\perp=18$~meV and $J^\perp=21$~meV. 
    }
    
\label{fig:beyond2lay}
\end{figure}

Our previous results for skyrmions in bilayers carry over to systems with more layers. For this purpose, we again use the effective parameter set to exclude exchange
frustration effects within the layers.
Energy barriers for skyrmions were obtained in three layer and four layer systems, with the magnetic atoms of the different layers all occupying the same lattice sites. Following our notation, these system are of the $\alpha\alpha$ type. We also studied a system with four layers and six layers with an $\alpha\beta$ stacking. For 
weak interlayer exchange, we calculated increased multiplicity of collapse mechanisms, in agreement with the bilayer results. Presenting this complexity is beyond the aim of this paper. We therefore present here only the regime of large interlayer exchange coupling. The energy barriers depending on the interlayer exchange $J^\perp$ of the skyrmions in the multilayer systems studied are shown in Fig.~\ref{fig:beyond2lay}(a) relative to the energy barrier of the skyrmion of the monolayer system.

As expected, the energy barriers for the three layer (four layer) skyrmions in the $\alpha\alpha$ system converge to three (four) times the energy barrier of the skyrmion in the monolayer system. However, it can be observed in Fig~\ref{fig:beyond2lay}(a) that a larger interlayer exchange coupling $J_C^\perp$ is needed in the case of the three and four layer system to force a simultaneous collapse of the skyrmions than in the bilayer. If we extrapolate the results obtained here for the skyrmions in the $\alpha\alpha$ systems (Fig~\ref{fig:beyond2lay}(b))
to a system with $L$ layers in which the atoms of all layers occupy the same lattice sites, we confirm the conjecture $\Delta E=L\Delta E_\text{mono}$ for the skyrmion in the multilayer system as long as $J^\perp> J_C^\perp$ holds. This is in agreement with the prediction in Ref.~\cite{Heil2019}.

It is the general view that an increase in magnetic material leads to an increase in the stability of skyrmions in magnetic multilayers. To ensure simultaneous behavior of these skyrmions, it is often concluded that the largest possible interlayer exchange is desirable. Our calculations for the $\alpha\alpha$ systems confirm this. If we move to the $\alpha\beta$ systems, which are relevant for real layered materials, we also find that increasing the number of layers increases the energy barrier of the skyrmions (Fig~\ref{fig:beyond2lay}(a))
consistent with the studies of Hoffmann \textit{et al.}\cite{hoffmann2020}. 

However, the situation is more complicated. What can be deduced from the data shown in Fig~\ref{fig:beyond2lay}(a)
is that the maximum stability for skyrmions in multilayers is achieved for a certain value of interlayer exchange. The maximum of the energy barrier for the skyrmion in the four layer $\alpha\beta$ system is below $3.5$ times the energy barrier of the skyrmion in the monolayer system and is obtained for $J^\perp\approx 6$~meV. The maximum achievable energy barrier for the skyrmion in the six layer $\alpha\beta$ system is even below $5$ times the energy barrier in the monolayer system. This is in contrast to the common belief that interlayer exchange coupling does not affect the stability of multilayer skyrmions as long as it is strong enough to allow simultaneous behavior of the skyrmion. 

Comparing the different $\alpha\beta$ systems also indicates that the decrease of the energy barrier for high interlayer exchange couplings occurs with a more negative slope the more layers are involved. This leads to the fact that the energy gain in terms of skyrmion stability by adding another layer decreases with increasing interlayer exchange coupling (Fig.~\ref{fig:beyond2lay}(b)). We propose that the energy barrier of skyrmions in fcc- or hcp-stacked multilayer systems with $L$ 
layers is thus given by 
$\Delta E= g(J^\perp)\cdot L \Delta E_\text{mono}$. Where the function $g(J^\perp)<1$ attributes to the fact that optimizing the skyrmion stability through adding more layers relies on the choice of the optimal interlayer exchange. This counterintuitive result provides an important contribution to the understanding of skyrmion stability in magnetic multilayers and is visualized in Fig.~\ref{fig:beyond2lay}(b). Here we extracted the energy barriers for fixed values of $J^\perp$ from Fig.~\ref{fig:beyond2lay}(a) and plotted versus the number of layers.

\subsection{Lifetime of bilayer skyrmions}
\label{ssec:lifetime}
In the preceding sections we discussed the dependence of the energy barrier on the interlayer exchange in magnetic bilayer systems, which is the dominant contribution to the lifetime at low temperatures due to the exponential term in Eq.~(\ref{Eq:arrhenius}). But, as reported in Ref.~\cite{malottki2019,varentcova2020}, the effect of the change of the pre-exponential factor should not be underestimated. Therefore, we present the calculation of the pre-exponential factor $\tau_0$ for the generic example of the bilayer stacks based on the effective parameter set as discussed in Sec.~\ref{ssec:stacks_effective_model}. For the purpose of underlining our results concerning the stability of skyrmions in the high interlayer exchange coupling regime we only discuss here the regime where one saddle point configuration appears. The description of collapses containing an intermediate minimum should be done with Master's equation and lies beyond the scope of this paper.
The diagonalization of the Hessian matrix for the bilayer skyrmion and the saddle point configuration gives us the eigenvalues of the initial bilayer skyrmion $\epsilon_{A,i}$ and the saddle point configuration $\epsilon_{\text{Sp},i}$. The determined eigenvalues allow the calculation of the prefactors following Eq.~(\ref{eq:prefactor_radial}).

\begin{figure}
\includegraphics[scale=1]{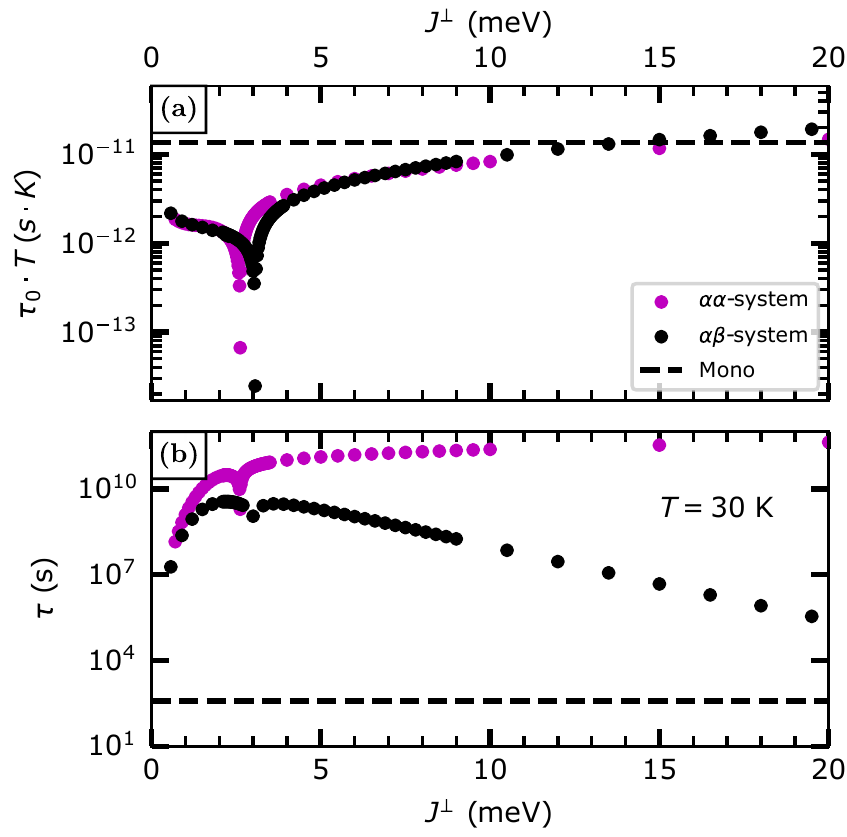}
\caption{(a) Temperature independent pre-exponential factor $\tau_0 \cdot T$ (Eq.~(\ref{gl:HTST})) for skyrmions in magnetic bilayer systems based on the magnetic monolayer system Pd/Fe/Ir(111) in $\alpha\alpha$- and $\alpha\beta$-stacking for various interlayer exchange couplings $J^\perp$. The dashed line indicates pre-exponential factor for skyrmions within the magnetic monolayer system Pd/Fe/Ir(111)\cite{malottki2019} for $B=4.0$~T. (b) Skyrmion lifetime $\tau$ for $T=30$ K calculated with the Arrhenius law (Eq.~(\ref{Eq:arrhenius})) using the energy barriers $\Delta E$ from Fig.~\ref{fig:eig_transition}~(a) and the prefactor $\tau_0$ displayed in part (a) of this figure.}
\label{fig:prefactor_lifetime}
\end{figure}

Fig.~\ref{fig:prefactor_lifetime}(a) shows a highly similar behavior for the $\alpha\alpha$ and $\alpha\beta$ stack regarding the pre-exponential factor $\tau_0$. A sharp decline of $\tau_0$ occurs around $J^\perp=3.0$~meV followed by an increase towards prefactor of the magnetic monolayer system, which is indicated by the dashed line. This narrow sink is produced by the softening of layer-aligning saddle point mode which approaches zero in this regime (See Fig.~\ref{fig:eig_transition}). The softening leads to a division by zero
in Eq.~(\ref{eq:prefactor_radial}) and therefore $\tau_0$ approaches zero for $J^\perp\approx J_C^\perp$. In this region the applicability of the harmonic approximation is questionable. Nevertheless, it is remarkable that the prefactor reduces the stability of the bilayer skyrmions for both stackings compared to the prefactor of the skyrmion in the magnetic monolayer system 
(dashed line in Fig.~\ref{fig:prefactor_lifetime}). We 
attribute this to an increased entropic difference between the transition state and the skyrmion state for intermediate interlayer exchange couplings as the number of possible transition mechanisms reduces with increased exchange couplings between the layers. 

In 2017 Wild \textit{et al.}\cite{wild2017} investigated the lifetime of skyrmions in B20-compounds. Changes in the magnetic field which lead to an increased energy barrier were counterbalanced by changes in the pre-exponential factor by 30 orders of magnitude leading to a substantial reduction of the lifetime of skyrmions by entropic effects\cite{wild2017}. However we expect that the increase in the energy barrier for skyrmions in systems with multiple magnetic layers always goes along with such a entropic induced decrease of the pre-exponential-factor $\tau_0$ for low $J^\perp.$ As the interlayer exchange coupling increases above $J^\perp\approx 15$~meV the prefactor of the bilayer systems reaches the prefactor for skyrmions in the monolayer system (Fig.~\ref{fig:prefactor_lifetime}(a)). Note, that the visualization in Fig.~\ref{fig:prefactor_lifetime}(a) is valid for all temperatures 
$T$ since the linear dependance in 
Eq.~(\ref{eq:prefactor_radial}) allows to display $\tau_0 \cdot T$.
Since the order of collapses does not matter for the SR collapses for $J^\perp<J_C^\perp$ two saddle points exist here and we multiplied $\tau_0^{-1}$ by a further factor of two in this regime.

In Fig.~\ref{fig:prefactor_lifetime}(b) we calculated the lifetime $\tau$ for the exemplary temperature $T=30$ K. For the shown parameter range 
of $J^\perp$ the stability of the bilayer skyrmion is always enhanced compared to the skyrmion in the magnetic monolayer system. The results of this section exemplify that the effects of changing the pre-exponential factor are relatively small when varying the interlayer exchange compared to the influence of the energy barrier on the lifetime of the bilayer skyrmions discussed here. 
Therefore, one can associate the results of the previous sections regarding the energy barriers of bilayer skyrmions directly with the stability of these skyrmions.

\section{Conclusion}\label{chap:conclusion}
In this work, we investigated fundamental properties of skyrmion stability in magnetic multilayer systems.
We considered multilayers built from single Fe layers with the magnetic properties taken from
the well-studied film system Pd/Fe/Ir(111) and coupled by interlayer exchange of variable
strength $J^\perp$.
The layers are either stacked in $\alpha\alpha$ order, in which the magnetic atoms are placed on top of each other, or in $\alpha\beta$ order, as it appears for fcc or hcp stacked systems. %
It turns out that for $\alpha\beta$-stacking, the interlayer exchange coupling acts as an exchange-bias to the system affecting the magnetic phase and skyrmion stability, while no such effect occurs for the $\alpha\alpha$ stacking.

For both stacking orders of magnetic bilayers, we found the expected simultaneous collapse of skyrmions in both Fe layers when $J^\perp$ exceeds a critical interlayer exchange, $J_C^\perp$. %
The collapse 
splits
into the successive annihilation of skyrmions in individual layers for small $J^\perp$, which can be seen as the bilayer analogue to the occurrence of chiral magnetic bobbers in bulk systems \cite{Rybakov2015}. %
For intermediate strengths of $J^\perp$, a rich phase space of collapse mechanisms arises, in which the interlayer exchange interaction can favor a mix of semi-successive chimera and radial symmetric mechanisms. %

Our analysis of the eigenvalue spectrum of the bilayer system revealed the layer-aligning eigenmode, which is responsible for the transition from the semi-successive radial (SSR) collapse to the simultaneous collapse. %
We found, that this transition can be described accurately by Landau's theory for continuous phase transitions, which provides a stable definition of the critical interlayer exchange $J_C^\perp$. %
This can help to design multilayer systems in the simultaneous collapse regime, which is desirable for most applications since the annihilation processes become more complex and thus harder to control for 
less strongly coupled systems. %

Harmonic transition state theory calculations show a small dependence of the prefactor of the interlayer exchange constant and the number of magnetic layers which indicates only a minor role of entropic effects in the investigated parameter space. %
However, the situation could be different for couplings below $J_C^\perp$ where the role of additional multilayer eigenmodes is more complex as well as for other systems, in which the exchange bias induced by interlayer exchange could lead to more drastic changes of the entropy at the skyrmion or saddle point state. %


As expected, the energy barriers of the $\alpha\alpha$-stacking order 
increase linear with the number of magnetic layers, L, as long as $J>J_C^\perp$.
The critical value $J_C^\perp$, on the other hand, is increasing with $L$ as well and is thus limiting the number of layers in the simultaneous collapse regime for a given $J^\perp$. %
These results 
are consistent
with 
the notion
that a strong interlayer exchange is desirable for increasing skyrmion stability. %

However, 
for an $\alpha\beta$ stacking, collinear structures are favored by the nearest-neighbor interlayer exchange interaction. %
This leads to the existence of a sweet spot for the strength of $J^\perp$, at which the increase of the total energy barrier due to the increased number of layers and the reduction of the energy barrier caused by strong interlayer coupling is optimized. %

It turns out, that the optimal choice of $J^\perp$ 
changes
drastically with the number of coupled magnetic layers, the other interaction parameters, and the crystal structure of the multilayer. %
These results occur systematically and consistent for our model systems with and without intralayer exchange frustration and over a large interval of interlayer exchange parameters. %
Therefore, we expect these effects to apply rather general and that they have to be taken into account in order to accurately predict skyrmion stability in magnetic multilayer systems.
In contrast to the common assumption that the total energy barrier of
skyrmion collapse in a multilayer scales as
$\Delta E=L\Delta E_\mathrm{Mono}$, we have demonstrated that it
is only an upper boundary 
and that the actual energy barrier can be much below this desired value. %

\begin{acknowledgments} We gratefully acknowledge financial
support from the Deutsche Forschungsgemeinschaft (DFG, German Research Foundation)
via project no.~414321830 (HE3292/11-1) and no.~418425860 (HE3292/13-1), the Icelandic Research Fund (Grant No. 217750 and 184949), and the Russian Science Foundation (Grant No. 19-72-10138).
\end{acknowledgments}

\appendix

\begin{figure}
\includegraphics[scale=1]{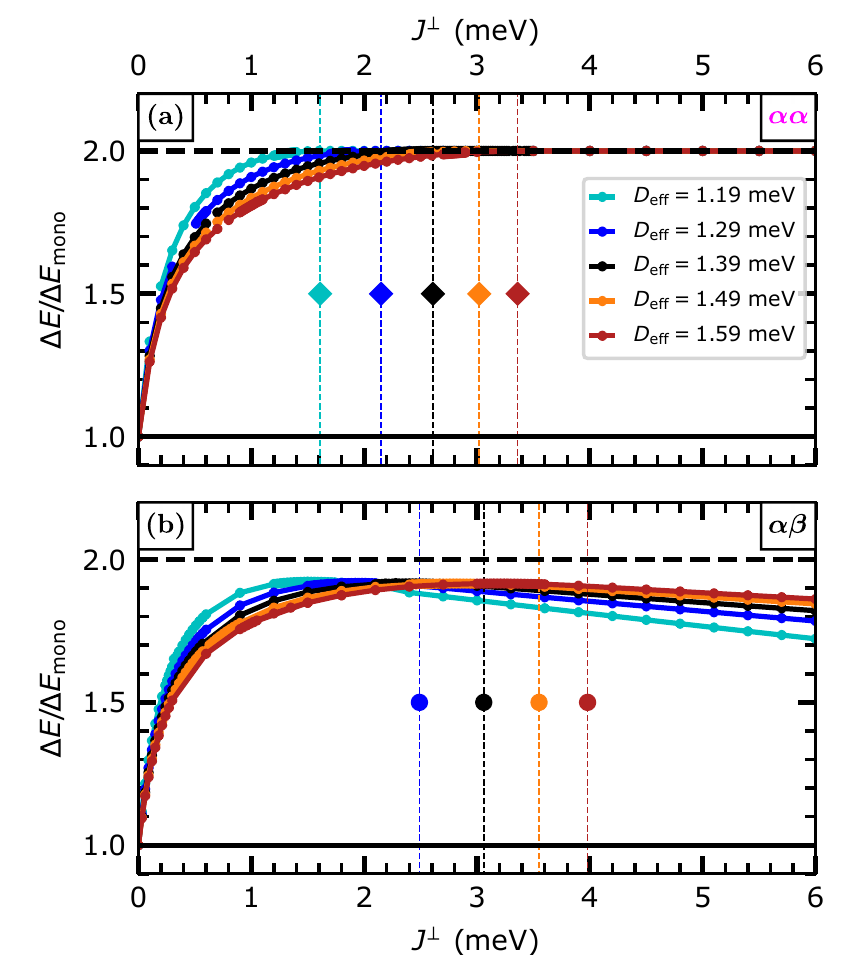}
\caption{(a,b) Energy barrier $\Delta E$ of bilayer skyrmions in the $\alpha\alpha$- and $\alpha\beta$-system for various interlayer exchange couplings $J^\perp$ relative to the energy barrier of a skyrmion in the underlying monolayer system $\Delta E_\text{mono}$. 
The variation of the DMI is indicated through the color code in the legend. (a) displays the $\alpha\alpha$- and (b) presents the $\alpha\beta$-system. The vertical dashed lines indicate the positions of $J_C^\perp$ (See Fig.~\ref{fig:eig_transition}). These critical interlayer exchange couplings $J_C^\perp$ are displayed in dependence of $\Delta E_\text{mono}$ in Fig.~\ref{fig:tune_barrier}.}
\label{fig:app_vary_mono}
\end{figure}

\section{Interaction constants for Pd/Fe/Ir(111)}
\label{sec:app_constants_pdfeir}
As the systems treated in this work are based on the DFT parametrized magnetic monolayer system fcc-Pd/Fe/Ir(111)\cite{malottki2017,malottki2019} the interaction constants of the corresponding extended Heisenberg model are listed in Tab.~\ref{tab: Pd_Fe_Ir_parameters}. 
\begin{table*}
\centering
\caption {Value of the $i$-th nearest neighbor intralayer exchange $J_i^{||}$ (meV), the Dzyaloshinskii-Moriya interaction constants $D_i$ (meV) and the magnetocrystalline anisotropy (MAE) $K$ (meV/Fe-atom) for the magnetic monolayer system fcc-Pd/Fe/Ir(111). These values originate from first-principles calculations from Ref.~\cite{malottki2017}. The positive $K>0$ parameters represent an out-of-plane easy axis for the anisotropy. }
\begin{ruledtabular}
\begin{tabular}{lccccccccccc}
model & $J_1^{\parallel}$& $J_2^{\parallel}$& $J_3^{\parallel}$& $J_4^{\parallel}$& $J_5^{\parallel}$& $J_6^{\parallel}$& $J_7^{\parallel}$& $J_8^{\parallel}$& $J_9^{\parallel}$& $D_1$& $K$ \\ \hline
NRE model & $14.40$ & $-2.48$ & $-2.69$ & $0.52$ & $0.74$ & $0.28$ & $0.16$ & $-0.57$ & $-0.21$ & $1.0$ & $0.7$ \\
eff. model & $3.68$ & - & - & - & - & - & - & - & - & $1.39$ & $0.7$
\end{tabular}
\end{ruledtabular}
\label{tab: Pd_Fe_Ir_parameters}
\end{table*}

\section{Identification of collapse mechanisms}
\label{sec:app_minmz}
The examination of bilayer skyrmion collapses as a function of interlayer exchange coupling $J^\perp$ in Sec.~\ref{ssec:introduction_to_mechanisms} revealed several mechanisms. In Fig.~\ref{fig:compare_stacks_summarize} in Sec.~\ref{ssec:barriers}, an overview of the parameter range of $J^\perp$ for the respective collapse mechanisms is indicated by the background color. While the boundaries between the mechanisms in the low interlayer exchange region are determined from the corresponding energy barriers in Fig.~\ref{fig:compare_stacks_summarize}, this section shows how the boundaries between the regimes in the intermediate and high coupling regions were determined. 

\begin{figure*}
\includegraphics[scale=1]{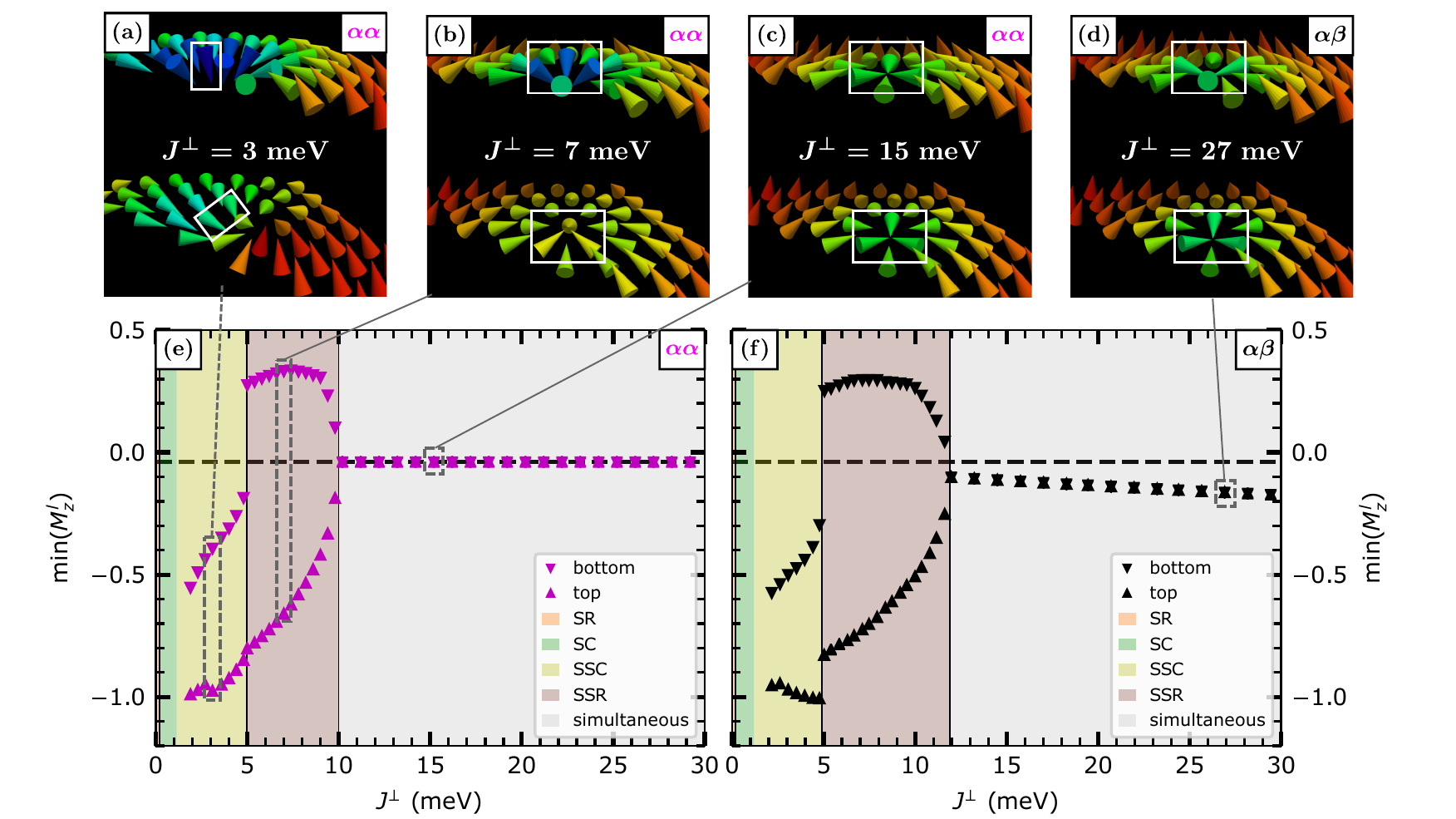}
\caption{(a-c) Representation of the saddle point configuration of the bilayer skyrmion collapse in the $\alpha\alpha$ system for $J^\perp=3$~meV, $J^\perp=7$~meV and $J^\perp=15$~meV. The $m_i^z$-component of the magnetic moments $i$ is emphasized by the color code. In each layer the magnetic moments $i$ with the minimum $m_i^z$-component are highlighted with a white box. (d) Analog representation for the saddle point configuration of a collapse in the $\alpha\beta$-system for $J^\perp=27$~meV. (e) Visualization of the minimum component of the magnetization in the z-direction for collapses of bilayer skyrmions in the $\alpha\alpha$-system for various $J^\perp$ for both layers $l=1,2$. The background color indicates the regime of the collapse types as introduced in Sec.~\ref{ssec:introduction_to_mechanisms}. (f) Analog visualization to (e) for the $\alpha\beta$-system. For comparison the corresponding value for the monolayer skyrmion collapse in Pd/Fe/Ir(111) is indicated as dashed black line in (e) and (f). All calculations where done with the NRE parameter set and only each third data point is shown for better visibility in (e),(f).}
\label{fig:app_minmz}
\end{figure*}

In Fig.~\ref{fig:app_minmz}(a-d), saddle point configurations are shown for different values of $J^\perp$ for the $\alpha\alpha$- and $\alpha\beta$-systems. While Fig.~\ref{fig:app_minmz}(a) represents a saddle point of semi-successive chimera (SSC) collapse, Fig.~\ref{fig:app_minmz}(b) shows semi-successive radial (SSR) collapse, and Fig.~\ref{fig:app_minmz}(c) demonstrates simultaneous saddle point in the $\alpha\alpha$-system. An example of a saddle point for large $J^\perp$ for the $\alpha\beta$-system is given in Fig.~\ref{fig:app_minmz}(d). If one searches for the magnetic moment with the minimum magnetization in the z-direction for each layer (see white boxes in Fig.~\ref{fig:app_minmz}(a-d)) and plots this value above $J^\perp$, a systematic classification of the mechanisms can be made. For the $\alpha\alpha$-system this representation can be found in Fig.\ref{fig:app_minmz}(e) and for the $\alpha\beta$-system in Fig.\ref{fig:app_minmz}(f).  At this point it is important to mention that the indexing of the layers is arbitrary, since the order of the skyrmion transitions in the different layers is not fixed. We now define the transition between the SSC regime to the SSR collapse mechanism by the jump visible in Fig.~\ref{fig:app_minmz}(e,f) for $J^\perp\approx 5$~meV. The transition from the SSR regime to the region of simultaneous collapse can again be defined by the point at which the minimum magnetization in the $z$-direction coincides in both layers.

\section{Varying the monolayer skyrmion barrier}
\label{sec:app_vary_mono}
Similar to Fig.~\ref{fig:eig_transition}(a) and (c) we varied the interlayer exchange coupling and calculated the energy barriers of bilayer skyrmions. In addition, we varied the energy barrier of the skyrmions in each layer by changing the value of the DMI ($D_\text{eff}$). This is described in Sec.~\ref{ssec:vary_monolayer_barrier}. The results of these calculations are shown in Fig.~\ref{fig:app_vary_mono}(a) for the $\alpha\alpha$ systems and in Fig.~\ref{fig:app_vary_mono}(b) for the $\alpha\beta$ systems. We calculated the critical interlayer exchange parameters $J_C^\perp$, which mark the onset of the regime of the simultaneous skyrmion collapse, by calculating the eigenvalue spectrum as presented in Fig.~\ref{fig:eig_transition}(b,d) and fitting of Eq.~(\ref{gl:landau_curvature}) to the eigenvalues of the layer-aligning mode above $J_C^\perp$.

\bibliographystyle{apsrev4-1}

\end{document}